\documentclass[superscriptaddress,
twocolumn,secnumarabic,nofootinbib]{revtex4-1}
\usepackage{amsmath}
\usepackage[utf8]{inputenc}
\usepackage[T1]{fontenc}

\pdfoutput=1

\usepackage{bm}
\usepackage{times}
\usepackage{braket}
\usepackage{color,graphicx}
\usepackage[utf8]{inputenc}
\usepackage[T1]{fontenc}
\usepackage[export]{adjustbox}

\usepackage{enumitem}

\usepackage{verbatim}

\usepackage{float}

\usepackage{colortbl}
\usepackage{longtable}
\usepackage{tabularx}

\usepackage{multirow}

\usepackage{xcolor}

\usepackage{lipsum}

\usepackage[normalem]{ulem}

\definecolor{LightGray}{gray}{0.96}
\definecolor{Gray}{gray}{0.94}
\definecolor{nicered}{rgb}{0.7,0.1,0.1}
\definecolor{nicegreen}{rgb}{0.1,0.5,0.1}

\usepackage[colorlinks,
            linkcolor=black,
            filecolor=black,
            anchorcolor=black,
            urlcolor=black,
            citecolor=blue,
            bookmarks=false,
            ]{hyperref}

\allowdisplaybreaks

\setlongtables

\begin{document}

\title{Leptogenesis due to oscillating Higgs field}

\author{Seishi Enomoto} \email[]{seishi@mail.sysu.edu.cn}
\author{Chengfeng Cai,} \email[]{caichf3@mail.sysu.edu.cn}
\author{Zhao-Huan Yu} \email[]{yuzhaoh5@mail.sysu.edu.cn}
\author{and Hong-Hao Zhang} \email[]{zhh98@mail.sysu.edu.cn}
\affiliation{School of Physics, Sun Yat-sen University, Guangzhou 510275, China}
  
\begin{abstract}
We propose a new leptogenesis scenario in which the lepton asymmetry and matter particles are simultaneously generated due to the coherent oscillating Higgs background.   To demonstrate the possibility of our scenario, we consider the type-I seesaw model as an illuminating example and show the numerical analysis.  In order to generate the required lepton number $|n_L/s| = 2.4 \times 10^{-10}$, we find that the scales of the Higgs background oscillation is required to be higher than $10^{14}$ GeV.
\end{abstract}
\pacs{XX.XX}
\maketitle

\section{Introduction}
The observation of the cosmic microwave background (CMB) supports two important cosmological events\cite{Aghanim:2018eyx,Ade:2015xua}.  One is the cosmic inflation that causes the exponential expansion of the Universe.  The evidence is shown in the scale invariance of the power spectrum.  Another important event is the big bang nucleosynthesis (BBN) in which the abundance of baryons (hydrogen, deuterium, helium, etc.) is fixed from a second to 3 minutes after inflation.  According to the theory of the BBN, the required initial conditions to realize the current Universe are a temperature higher than a few MeV and the baryon number to photons number ratio $\eta = n_B/n_\gamma\sim 6\times 10^{-10}$.  Despite the CMB information is originated after 380,000 years later from the BBN, the observation shows $\eta=(6.10\pm0.04)\times10^{-10}$\cite{Ade:2015xua,Cyburt:2015mya}.  However, these two events cannot connect directly because the temperature of the Universe after inflation might go to zero due to the extreme dilution.  On the other hand, the BBN must start with a high-temperature scale at least more than a few MeV.  Therefore, the Universe must be heated due to some mechanism after inflation.

The (re)heating theory has been developed by many authors (for review, see e.g. \cite{Kofman:1997yn,Amin:2014eta,Lozanov:2019jxc,Enomoto:2020epjc}).  Typically it is described by the decay of the inflation field into other particles after inflation, but a picture of the perturbative decay is not correct.  Taking into account a particle coupled to the oscillating inflation field, the non-perturbative particle production occurs and the produced particle number can grow exponentially by parametric resonance\cite{Dolgov:1989us,Traschen:1990sw,Kofman:1994rk,Kofman:1997yn}.  This process happens before the Universe is thermalized, so-called preheating era.

On the other hand, inflation makes our Universe no particle state because any particles are extremely diluted.  Thus, the Universe must evolve from no baryon number state to non-zero baryon number state.  This scenario is called baryogenesis that requires a small asymmetry between baryons and anti-baryons.  The Standard Model (SM) cannot produce enough baryon asymmetry from a symmetric Universe.
At present, one of the hopeful scenarios is the thermal leptogenesis \cite{Fukugita:1986hr} in which superheavy right-handed neutrinos and their interactions are added to the SM.  The decay of right-handed neutrinos can generate lepton number.  This generated lepton number can be converted into the baryon number through the sphaleron process \cite{Harvey:1990qw} after the decay of the right-handed neutrinos.

Several baryogenesis scenarios associated with preheating have been investigated in the past.  The non-perturbative particle production due to the oscillating inflation field allows creating heavier particles than the inflaton mass, while the ordinary reheating theory cannot kinematically.  This fact can be applied to the baryogenesis scenario in which the parent particles are heavier than the oscillating field\cite{Kolb:1996jt,Kolb:1998he,Giudice:1999fb}.  The oscillating background field that couples to the baryon or the lepton can also induce the chemical potentials to the baryon or the lepton numbers.  The scenario due to such chemical potential has been discussed in \cite{Dolgov:1996qq,Kusenko:2014uta}.

In this paper we propose a new scenario of the baryogenesis that occurs in the preheating era.  In contrast to the scenarios mentioned above, our scenario generates the baryons or leptons and their asymmetry simultaneously due to an oscillating background field.  In earlier studies, it was pointed out the possibility of the asymmetry production in the preheating era with simple scalar cases that have charge violating interactions and C- and CP-violating parameters\cite{Funakubo:2000us,Rangarajan:2001yu,Enomoto:2017rvc}.  But it is still not clear that there exist realistic models that can generate enough baryon or lepton asymmetry.

To show the story with a concrete model, we consider the type-I seesaw model as an example. 
Then our scenario corresponds to a new-type of non-thermal leptogenesis.  A sketch of our scenario is that the lepton asymmetry is produced by the non-perturbative particle production of the left-handed neutrinos due to the coherently oscillating Higgs background.  For example, let us consider a lepton number violating operator $C^{AB}(H\ell^A)(H\ell^B)$ first proposed by Weinberg\cite{Weinberg:1979sa}, where $C^{AB}$ is a coupling and $H$ is the Higgs doublet, and $\ell^A$ is the lepton doublet in the generation $A$.  This interaction also represents mass terms of the left-handed neutrinos in the case that the Higgs has a vacuum expectation value.  If the Higgs vacuum expectation value varies non-adiabatically, the left-handed neutrinos would be produced.  An important point is that this neutrino production violates the lepton number.  Therefore, if the theory has C- and CP-violating parameters, then the lepton number asymmetry is also generated at the same moment.  However, this process must happen at a lower energy scale than the scale of the coupling $C^{AB}$.  Otherwise, the backreactions or the neutrino-Higgs scatterings happen, erasing the generated asymmetry.  Similar situations are investigated\cite{Pearce:2015nga,Pascoli:2016gkf,Turner:2018mwh,Pascoli:2018cqk,Wu:2019ohx,Lee:2020yaj}, but the interaction contents or the generation structure in those studies are different from ours.  The models in \cite{Kusenko:2014lra,Pearce:2015nga,Wu:2019ohx} were considered the type-I seesaw and a dimension six operator that is proportional to $B+L$ current.  Then, the lepton asymmetry can be generated by a single generation.  In \cite{Pascoli:2016gkf,Turner:2018mwh,Pascoli:2018cqk}, the case that the coupling of the Wienberg operator is time-dependent was studied.  Compared with those studies, our scenario is considered the type-I seesaw with three generations of the left- and the right-handed neutrinos, without any other higher dimensional operator and without the time-dependent couplings.  Recently, the similar scenario to ours have been discussed in \cite{Lee:2020yaj} in which the flavour of the left-handed neutrino is single.

Since it is difficult to obtain the analytic behavior, we will demonstrate the formulation and show the results by the numerical calculation.  
For simplicity, we neglect the spatial expanding effect in the later calculation.  Furthermore, we assume that the energy scale of the produced left-handed neutrinos is much lower than the mass scale of the heavy right-handed neutrinos in order to avoid backreactions that would erase the generated asymmetry.  Then the production of the right-handed neutrinos associated with the Higgs is forbidden, and the scattering of the left-handed neutrinos and the Higgs intermediated with the heavy right-handed neutrinos is also suppressed.

The paper is organized as follows.  In section \ref{sec:formulation_of_op_eq},  we derive the basic operator equations and construct the effective theory in which the right-handed neutrinos do not appear.  Using these results, we construct the equations of motion for two-point functions and the Higgs background to evaluate the lepton asymmetry in section \ref{sec:ana_lepton_asym}.  The numerical results are also shown in this section.  We summarize our conclusion and discuss in section \ref{sec:conclusion}.

\section{Formulation of operator equations} \label{sec:formulation_of_op_eq}
The goal of this paper is to demonstrate that the lepton asymmetry can be generated by the oscillating Higgs background.  At first, we explain what we must calculate to know the generated lepton asymmetry.
As mentioned in the previous section, we consider our model with the type-I seesaw model that includes the SM and the three generations of the right-handed neutrinos.  Furthermore, we neglect the spacial expanding effect in the later calculation for simplicity and assume that the energy scale of the produced neutrinos is much lower than the heavy right-handed neutrino scale in order to avoid backreactions erasing the generated asymmetry.  In our formulation, we use the notation of the metric as $g_{\mu\nu}={\rm diag} (+1,-1,-1,-1)$, and use the two-component spinors as the representation of fermions.

The net lepton number can be defined by a vacuum expectation value of $U(1)$ Noether charge of the leptons as
\begin{eqnarray}
 L &\equiv& \int d^3x \sum_A \frac{1}{2}
  \left( (\langle \nu_L^{A\dagger}\bar{\sigma}^0\nu_L^A\rangle-\langle \nu_L^A\sigma^0\nu_L^{A\dagger}\rangle) \right. \nonumber \\
   & & \qquad \qquad \qquad + (\langle e_L^{A\dagger}\bar{\sigma}^0e_L^A \rangle - \langle e_L^A\sigma^0e_L^{A\dagger} \rangle) \nonumber \\
   & & \left. \qquad \qquad \qquad
   - (\langle e_R^{cA\dagger}\bar{\sigma}^0 e_R^{cA}\rangle - \langle e_R^{cA}\sigma^0 e_R^{cA\dagger}\rangle) \right) \label{eq:lepton_number1}
\end{eqnarray}
where the superscript $A$ runs the generation, and $\nu_L$, $e_L$ and $e_R^c$ are the left-handed neutrino, the left-handed electron and the charge conjugate of the right-handed electron, respectively.  The lepton number density $n_L$ can be obtained by $n_L=L/V$ where $V\equiv \int d^3x$ is a spatial volume of the system.
As shown in (\ref{eq:lepton_number1}), we need to follow every leptonic two-point function.
To derive the equations of motion for each two point function, at first we derive the operator equations for leptons.

In later calculation, we derive the equations of motion for each operator field and construct the differential equations for all the required two-point functions using the operator equations.  Then, we solve them numerically, and we follow the time evolution of the generated lepton asymmetry referring to the solved two-point functions.

\subsection{Lagrangian}
The Lagrangian relating to the lepton sector is given by
\begin{eqnarray}
 \mathcal{L}_{\rm lepton}
  &=& \sum_A\left(\ell^{aA\dagger}\bar{\sigma}^\mu iD_\mu\ell^{aA} + e_R^{cA\dagger}\bar{\sigma}^\mu iD_\mu e_R^{cA} \right. \nonumber \\
  & & \qquad \left. + \nu_R^{cA\dagger}\bar{\sigma}^\mu i\partial_\mu\nu_R^{cA} \right) \nonumber\\
  & & -\sum_{A,B}\left(\frac{1}{2}M_R^{AB}\nu_R^{cA}\nu_R^{cB}+\sqrt{2}y_e^{AB}H^{a\dagger}\ell^{aA}e_R^{cB} \right. \nonumber \\
  & & \qquad \quad \left. - \sqrt{2}y_\nu^{AB}\epsilon^{ab}H^a\ell^{bA}\nu_R^{cB}+({\rm h.c.})\right)
\end{eqnarray}
where superscripts $A, B$ run 3 generations (flavor basis) and $a, b$ run $SU(2)_L$ components, and mass matrix of the right-handed neutrinos $M_R$ is chosen to be diagonal and real components.
We choose the unitary gauge as $H^1=0, \: H^2=h/\sqrt{2}, \: \ell^1=\nu_L, \: \ell^2=e_L$,
then the Lagrangian by each component is shown by
\begin{eqnarray}
 \mathcal{L}_{\rm lepton}
  &=& \sum_A\left( \nu_L^{A\dagger}\bar{\sigma}^\mu i\partial_\mu\nu_L^A+e_L^{A\dagger}\bar{\sigma}^\mu i\partial_\mu e_L^A \right. \nonumber \\
  & & \qquad \left. + e_R^{cA\dagger}\bar{\sigma}^\mu i\partial_\mu e_R^{cA} +\nu_R^{cA\dagger}\bar{\sigma}^\mu i\partial_\mu\nu_R^{cA}\right. \nonumber \\
  & & \qquad+ g_YB_\mu\cdot\left(-\frac{1}{2}\nu_L^{A\dagger}\bar{\sigma}^\mu \nu_L^A
   -\frac{1}{2}e_L^{A\dagger}\bar{\sigma}^\mu e_L^A \right. \nonumber \\
  & & \qquad \qquad \qquad \quad \left. +e_R^{cA\dagger}\bar{\sigma}^\mu e_R^{cA}\right)\nonumber\\
  & & \qquad +\frac{1}{2}g_WW_\mu^3\cdot\left(\nu_L^{A\dagger}\bar{\sigma}^\mu\nu_L^A-e_L^{A\dagger}\bar{\sigma}^\mu e_L^A\right)\nonumber\\
  & & \qquad +\frac{1}{2}g_W(W_\mu^1-iW_\mu^2)\cdot\nu_L^{A\dagger}\bar{\sigma}^\mu e_L^A \nonumber \\
  & & \qquad \left.+ \frac{1}{2}g_W(W_\mu^1+iW_\mu^2)\cdot e_L^\mu\bar{\sigma}^\mu \nu_L^A\right) \nonumber\\ 
  & & -\sum_{A,B}\left(\frac{1}{2}M_R^{AB}\nu_R^{cA}\nu_R^{cB}+y_e^{AB}he_L^Ae_R^{cB}\right. \nonumber \\
  & & \qquad \qquad \left. +y_\nu^{AB}h\nu_L^A\nu_R^{cB}+({\rm h.c.})\right) \label{eq:lagrangian_component}
\end{eqnarray}
where $B_\mu$ and $W_\mu^a$ for $a=1,2,3$ are the gauge boson fields, $g_Y$ and $g_W$ are the gauge couplings corresponding to $U(1)_Y$ and $SU(2)_L$ gauge symmetries, respectively.

\subsection{Equations of motion}
\subsubsection{Approximate solution for right-handed neutrinos}
At first, we derive the equations of motion for the right-handed neutrinos.  From (\ref{eq:lagrangian_component}), the operator equations for the right-handed neutrinos can be obtained as
\begin{eqnarray}
 0 &=& \bar{\sigma}^\mu\cdot i\partial_\mu \nu_R^{cA}-\sum_B\left(M_R^{AB}\nu_R^{cB\dagger}
  +(y_\nu^\dagger)^{AB}h\nu_L^{B\dagger}\right),  \label{eq:eom_rhn01} \\
 0 &=& \sigma^\mu\cdot i\partial_\mu \nu_R^{cA\dagger}-\sum_B\left(M_R^{AB}\nu_R^{cB} + (y_\nu^T)^{AB}h\nu_L^B \right).\label{eq:eom_rhn02}
\end{eqnarray}
Since we assume that the scale of the non-perturbative particle production is enough lower than the right-handed neutrino mass scale, we can construct an effective equation in which the right-handed neutrinos do not appear.  Assuming $M_R^{-1}\partial \ll 1$ in (\ref{eq:eom_rhn01}) and (\ref{eq:eom_rhn02}), we can obtain the approximate solution as
\begin{eqnarray}
 \nu_R^{cA\dagger}
  &=& \sum_B\left(-(M_R^{-1}y_\nu^\dagger)^{AB}h\nu_L^{B\dagger}+(M_R^{-1})^{AB}\bar{\sigma}^\mu\cdot i\partial_\mu \nu_R^{cB}\right) \nonumber \\
  &=& -\sum_B\left( (M_R^{-1}y_\nu^\dagger)^{AB}h\nu_L^{B\dagger}+ (M_R^{-2}y_\nu^T)^{AB}i\partial_\mu h\cdot \bar{\sigma}^\mu\nu_L^B \right.\nonumber \\
  & & \qquad \quad\left.+(M_R^{-2}y_\nu^T)^{AB}h\cdot \bar{\sigma}^\mu i\partial_\mu\nu_L^B \right) \nonumber \\
 & & + \mathcal{O}((M_R^{-1}\partial)^2). \label{eq:app_sol_rhn}
\end{eqnarray}
Note that the first term in the above equation is of the leading order, and the second and third terms are of the next-to-leading order.

\subsubsection{Left-handed neutrinos}
The operator equation for the left-handed neutrinos is given by
\begin{eqnarray}
 0 &=& \bar{\sigma}^\mu\cdot i\partial_\mu\nu_L^A-\frac{1}{2}(g_YB_\mu-g_WW_\mu^3)\cdot \bar{\sigma}^\mu\nu_L^A \nonumber \\
  & & +\frac{1}{2}g_W(W_\mu^1-iW_\mu^2)\cdot \bar{\sigma}^\mu e_L^A-\sum_B y_\nu^{*AB}h\nu_R^{cB\dagger}.
\end{eqnarray}
Substituting the approximate solution (\ref{eq:app_sol_rhn}) to the above equation, we obtain an approximate equation without the right-handed neutrinos as
\begin{eqnarray}
 0 &=& \bar{\sigma}^\mu\cdot i\partial_\mu\nu_L^A-\frac{1}{2}(g_YB_\mu-g_WW_\mu^3)\cdot \bar{\sigma}^\mu\nu_L^A \nonumber \\
  & & +\frac{1}{2}g_W(W_\mu^1-iW_\mu^2)\cdot \bar{\sigma}^\mu e_L^A + \sum_B(y_\nu^*M_R^{-1}y_\nu^\dagger)^{AB}h^2\nu_L^{B\dagger} \nonumber \\
  & & +\sum_B(y_\nu^*M_R^{-2}y_\nu^T)^{AB}\nonumber \\
  & & \qquad \times \left(ih\cdot \partial_\mu h+h^2\left(\frac{1}{2}g_YB_\mu -\frac{1}{2}g_WW_\mu^3\right)\right)\bar{\sigma}^\mu\nu_L^B \nonumber \\
  & & -\sum_B(y_\nu^*M_R^{-2}y_\nu^T)^{AB}h^2\cdot\frac{1}{2}g_W(W_\mu^1-iW_\mu^2) \bar{\sigma}^\mu e_L^B + \cdot\cdots, \label{eq:app_sol_lhn0}
\end{eqnarray}
where $\cdots$ means higher order terms.
In the above operator equation, we impose the following approximations:
\begin{equation}
 h \sim \langle h(t)\rangle, \qquad B_\mu \sim 0, \qquad W_\mu^a \sim 0. \label{eq:app_h_gauge}
\end{equation}
These indicate that the Higgs has a homogeneous background but the gauge fields do not have backgrounds\footnote{Although there are no backgrounds of gauge fields (one-point functions), two-point functions that correspond to their number density are not negligible.  In our scenario, the gauge bosons can be produced by the coherent oscillation of the Higgs background.  We will see in section \ref{sec:EoHB} that the bosonic two-point functions appear in the Higgs sector.}.
Then (\ref{eq:app_sol_lhn0}) becomes a simpler equation as
\begin{eqnarray}
 0 &=& \bar{\sigma}^\mu\cdot i\partial_\mu\nu_L^A\nonumber \\
  & & + \sum_B\left( (y_\nu^*M_R^{-1}y_\nu^\dagger)^{AB}\langle h \rangle^2\nu_L^{B\dagger} \right. \nonumber \\
  & & \qquad \quad \left. + i(y_\nu^*M_R^{-2}y_\nu^T)^{AB}\langle h\rangle \langle \dot{h}\rangle \cdot \bar{\sigma}^0\nu_L^B \right) + \cdots. \label{eq:eom_lhnu}
\end{eqnarray}

In (\ref{eq:eom_lhnu}), a matrix $y_\nu M_R^{-1}y_\nu^T\cdot\langle h\rangle^2$
means the masses of the left-handed neutrinos but it is not diagonalized in general.
This mass matrix can be diagonalized by Pontecorvo-Maki-Nakagawa-Sakata matrix $U_{\rm PMNS}$ as
\begin{eqnarray}
 m_\nu(t) &\equiv& -U_{\rm PMNS}^Ty_\nu M_R^{-1}y_\nu^TU_{\rm PMNS}\cdot\langle h(t)\rangle^2 \label{eq:diag_nu}
\end{eqnarray}
where $m_\nu$ is a diagonal matrix and every component is real.
Since $U_{\rm PMNS}^Ty_\nu M_R^{-1}y_\nu^TU_{\rm PMNS}$ is a constant matrix, it can be represented by the present Higgs vacuum expectation value and the present left-handed neutrino masses:
\begin{equation}
 U_{\rm PMNS}^Ty_\nu M_R^{-1}y_\nu^TU_{\rm PMNS}
  = -\frac{1}{\langle h_{\rm now}\rangle^2}m_{\nu,{\rm now}} \label{eq:diag_nu2}
\end{equation}
where
\begin{equation}
 \langle h_{\rm now}\rangle=246 \:{\rm GeV}, \qquad m_{\nu,{\rm now}}=\left(\begin{array}{ccc}m_1&&\\&m_2&\\&&m_3\end{array}\right),
\end{equation}
and $m_1, m_2, m_3$ are the masses of the left-handed neutrinos in mass basis.
Equation (\ref{eq:diag_nu2}) can also be rewritten as \cite{Casas:2001sr}
\begin{eqnarray}
 1 &=& -m_{\nu,{\rm now}}^{-1/2}U_{\rm PMNS}^Ty_\nu
  M_R^{-1}y_\nu^TU_{\rm PMNS}m_{\nu,{\rm now}}^{-1/2}\cdot\langle h_{\rm now}\rangle^2 \nonumber \\
   &=& -\left[m_{\nu,{\rm now}}^{-1/2}U_{\rm PMNS}^Ty_\nu M_R^{-1/2}\right] \nonumber \\
   & & \quad \times
  \left[M_R^{-1/2}y_\nu^TU_{\rm PMNS}m_{\nu,{\rm now}}^{-1/2}\right]\cdot\langle h_{\rm now}\rangle^2 \nonumber \\
  &=& O^TO \label{eq:diag_nu3}
\end{eqnarray}
where
\begin{equation}
 O\equiv iM_R^{-1/2}y_\nu^TU_{\rm PMNS}m_{\nu,{\rm now}}^{-1/2}\cdot\langle h_{\rm now}\rangle
\end{equation}
is an orthogonal complex matrix.  Using this matrix, the Yukawa matrix $y_\nu$ can be represented by
\begin{equation}
 y_\nu^T=-\frac{i}{\langle h_{\rm now}\rangle} M_R^{1/2}Om_{\nu,{\rm now}}^{1/2}U_{\rm PMNS}^\dagger.
\end{equation}
Hence, we can obtain
\begin{equation}
 y_\nu^*M_R^{-2}y_\nu^T = \frac{1}{\langle h_{\rm now}\rangle^2}
  U_{\rm PMNS}m_{\nu,{\rm now}}^{1/2}O^\dagger M_R^{-1}Om_{\nu,{\rm now}}^{1/2}U_{\rm PMNS}^\dagger.
\end{equation}

Finally, the operator equation (\ref{eq:eom_lhnu}) written by the mass eigenstate can be shown as
\begin{eqnarray}
 0 &=& \bar{\sigma}^\mu i\partial_\mu \nu_L^I +\sum_J\left(-[m_\nu(t)]^{IJ}\nu_L^{J\dagger}+i[Z(t)]^{IJ}\bar{\sigma}^0\nu_L^J\right) +\cdots \nonumber \\ \label{eq:eom_diag_nu}
\end{eqnarray}
where the superscripts $I,J$ run mass eigenstate indices,
\begin{equation}
 \nu_L^I \equiv \sum_A(U_{\rm PMNS}^\dagger)^{IA}\nu_L^A
\end{equation}
is a mass eigenstate of the left-handed neutrinos, and
\begin{eqnarray}
 \left[m_\nu(t)\right]^{IJ} &\equiv& \frac{\langle h(t)\rangle^2}{\langle h_{\rm now}\rangle^2} \cdot [m_{\nu,{\rm now}}]^{IJ} \\
 \left[Z(t)\right]^{IJ} &\equiv& \frac{\langle h(t)\rangle\langle \dot{h}(t)\rangle}{\langle h_{\rm now}\rangle^2}
  \cdot [m_{\nu,{\rm now}}^{1/2}O^\dagger M_R^{-1}Om_{\nu,{\rm now}}^{1/2}]^{IJ}. \nonumber \\
\end{eqnarray}
Note that $Z$ is a non-diagonal Hermitian matrix in general.
The CP-violating parameters are included by the non-diagonal complex elements.

Since it is convenient to treat the equations by the Fourier transformed representation, we represent (\ref{eq:eom_diag_nu}) by each Fourier mode.  We expand $\nu_L(x)$ by a plane wave and each helicity mode as
\begin{equation}
 [\nu_L(x)_\alpha]^I=\int\frac{d^3k}{(2\pi)^3}e^{i\mathbf{k\cdot x}}\sum_{s=\pm}(e_{\mathbf{k}}^s)_\alpha[\nu_\mathbf{k}^s(t)]^I
\end{equation}
where $(e_{\mathbf{k}}^s)_\alpha$ is an eigen-spinor for the helicity operator that satisfies\footnote{
In this paper, we use the following representation:
\begin{equation}
 (e_{\mathbf{k}}^s)_1 = \sqrt{\frac{1}{2}\left(1 + \frac{sk^3}{|\mathbf{k}|}\right)},
  \qquad (e_{\mathbf{k}}^s)_2 = se^{i\theta_{\mathbf{k}}} \sqrt{\frac{1}{2}\left(1 - \frac{sk^3}{|\mathbf{k}|}\right)}
\end{equation}
where
\begin{equation}
 e^{i\theta_{\mathbf{k}}} \equiv \frac{k^1 + ik^2}{\sqrt{(k^1)^2 + (k^2)^2}}
\end{equation}
is a phase defined by the $x$- and $y$-components of the momentum.
They satisfy the following orthogonality conditions;
\begin{equation}
 e_{\mathbf{k}}^{s \dagger} \bar{\sigma}^0 e_{\mathbf{k}}^r =
  e_{\mathbf{k}}^r \sigma^0 e_{\mathbf{k}}^{s \dagger} = \delta^{sr}, \qquad
 e_{\mathbf{k}}^s e_{-\mathbf{k}}^r = s e^{i\theta_{\mathbf{k}}} \delta^{sr}.
\end{equation}
}:
\begin{equation}
 k^i(\bar{\sigma}^ie_{\mathbf{k}}^s)^{\dot{\alpha}}=-s|\mathbf{k}|(\bar{\sigma}^0e_{\mathbf{k}}^s)^{\dot{\alpha}} \qquad (\:s=+\:\:{\rm or}\:\:-\:).
\end{equation}
Then (\ref{eq:eom_diag_nu}) in the Fourier space can be represented by
\begin{eqnarray}
 \partial_t \nu_\mathbf{k}^{sI}
  &=& is|\mathbf{k}|\nu_\mathbf{k}^{sI}
   +\sum_J\left(im_\nu^{IJ}\cdot se^{-i\theta_{\mathbf{k}}} \nu_{-\mathbf{k}}^{sJ\dagger} -Z^{IJ}\nu_\mathbf{k}^{sJ}\right)+\cdots. \label{eq:eom_lhn_ft}
\end{eqnarray}
Later, we will use this differential equation to derive each two-point functions.

\subsubsection{Left- and right-handed electrons}
Since the electron sector does not have interaction with right-handed neutrinos,
each equation for electrons is the same as in the SM.
Moreover, as long as we use the approximation (\ref{eq:app_h_gauge}),
the electron sector is separated from the left-handed neutrinos.
Therefore, even if the lepton asymmetry is generated in the neutrino sector, the effect does not influence the electron sector.
For this reason, we ignore the electron sector in this work.

\section{Analysis of lepton asymmetry} \label{sec:ana_lepton_asym}
As shown in the previous section, the net lepton number is defined by (\ref{eq:lepton_number1}).
We expect that the lepton asymmetry from the electron sector can be neglected.
Thus, we only consider the neutrino sector and obtain
\begin{eqnarray}
 n_L &\simeq& \frac{1}{V}\int d^3x \sum_A \frac{1}{2}\left(
    \langle \nu_L^{A\dagger}\bar{\sigma}^0\nu_L^A\rangle - \langle \nu_L^A\sigma^0\nu_L^{A\dagger}\rangle\right) \nonumber \\
  &=& \frac{1}{V}\int\frac{d^3k}{(2\pi)^3}\sum_I\sum_{s=\pm}\frac{1}{2}\left(
   \langle \nu_\mathbf{k}^{sI\dagger}\nu_\mathbf{k}^{sI}\rangle
   -\langle \nu_\mathbf{-k}^{sI}\nu_\mathbf{-k}^{sI\dagger}\rangle\right). \label{eq:lepton_num_rep}
\end{eqnarray}
Therefore, we need to follow the values of each two-point functions in order to know the net lepton number.

According to the CMB observation\cite{Aghanim:2018eyx}, the baryon number in the Universe is given by
\begin{equation}
 \left.\frac{n_B}{s}\right|_{\rm obs} = 8.6 \times 10^{-11}
\end{equation}
where $n_B$ is the baryon density and $s$ is the entropy density.
The lepton number required for achieving this observation can be estimated as follows.  After the lepton number is generated in our scenario, a part of it converts to the baryon number through the sphaleron process which is in equilibrium at the temperature $T \lesssim 10^{12}$ GeV.
The amount can be estimated by \cite{Harvey:1990qw} 
\begin{equation}
 \frac{n_B}{s} = -\frac{28}{79}\frac{n_L}{s}.
\end{equation}
Thus the required lepton number is
\begin{equation}
 \frac{n_L}{s} = -\frac{79}{28}\left.\frac{n_B}{s}\right|_{\rm obs} = - 2.4\times 10^{-10}. \label{eq:required_lepton}
\end{equation}

\subsection{Differential equations for two-point functions}
We can construct differential equations for the two-point functions using the operator equation (\ref{eq:eom_lhn_ft}).  The relevant differential equations are given by
\begin{eqnarray}
 \partial_t\langle\nu_\mathbf{k}^{sI\dagger}\nu_\mathbf{k}^{sJ}\rangle
  &=& -\sum_K\left(\langle\nu_\mathbf{k}^{sI\dagger}\nu_\mathbf{k}^{sK}\rangle(Z^*)^{KJ} \right. \nonumber \\
  & & \qquad \qquad \left.
   +(Z^*)^{IK}\langle\nu_\mathbf{k}^{sK\dagger}\nu_\mathbf{k}^{sJ}\rangle \right) \nonumber \\
  & & +im_\nu^{II}[se^{i\theta_\mathbf{k}}\langle\nu_\mathbf{-k}^{sI}\nu_\mathbf{k}^{sJ}\rangle] \nonumber \\
  & & -im_\nu^{JJ}[se^{i\theta_\mathbf{k}}\langle\nu_\mathbf{-k}^{sJ}\nu_\mathbf{k}^{sI}\rangle]^* \label{eq:eom_nu_bar_nu}\\
 \partial_t\langle\nu_\mathbf{-k}^{sI}\nu_\mathbf{-k}^{sJ\dagger}\rangle
  &=& -\sum_K\left(\langle\nu_\mathbf{-k}^{sI}\nu_\mathbf{-k}^{sK\dagger}\rangle Z^{KJ} \right. \nonumber \\
  & & \qquad \qquad \left.
   +Z^{IK}\langle\nu_\mathbf{-k}^{sK}\nu_\mathbf{-k}^{sJ\dagger}\rangle \right) \nonumber \\
  & & +im_\nu^{II}[se^{i\theta_\mathbf{k}}\langle\nu_\mathbf{-k}^{sJ}\nu_\mathbf{k}^{sI}\rangle]^* \nonumber \\
  & & -im_\nu^{JJ}[se^{i\theta_\mathbf{k}}\langle\nu_\mathbf{-k}^{sI}\nu_\mathbf{k}^{sJ}\rangle] \label{eq:eom_nu_nu_bar} \\
 \partial_t[se^{i\theta_\mathbf{k}}\langle\nu_\mathbf{-k}^{sI}\nu_\mathbf{k}^{sJ}\rangle]
  &=& 2is|\mathbf{k}|[se^{i\theta_\mathbf{k}}\langle\nu_\mathbf{-k}^{sI}\nu_\mathbf{k}^{sJ}\rangle] \nonumber \\
  & &-\sum_K\left([se^{i\theta_\mathbf{k}}\langle\nu_\mathbf{-k}^{sI}\nu_\mathbf{k}^{sK}\rangle](Z^*)^{KJ} \right. \nonumber \\
  & & \qquad \qquad \left.
   +(Z^*)^{IK}[se^{i\theta_\mathbf{k}}\langle\nu_\mathbf{-k}^{sK}\nu_\mathbf{k}^{sJ}\rangle]\right) \nonumber \\
  & & +im_\nu^{II}\langle\nu_\mathbf{k}^{sI\dagger}\nu_\mathbf{k}^{sJ}\rangle
   -im_\nu^{JJ}\langle\nu_{-\mathbf{k}}^{sI}\nu_\mathbf{-k}^{sJ\dagger}\rangle. \nonumber \\ \label{eq:eom_nu_nu}
\end{eqnarray}
Note that correlation functions of three or more points that provide the interaction effects do not appear in the above equations because we approximate the bosonic operators by background fields in the previous section.
Although it is still difficult to analytically solve these differential equations because of the time-dependent matrices $m_\nu$ and $Z$, the numerical analysis can be performed in a relatively simple way.  The initial conditions for each two-point function are chosen to be\footnote{Actually, the asymptotic solution of (\ref{eq:eom_lhn_ft}) corresponding to zero particle state is given by
\begin{equation}
 \nu_\mathbf{k}^{sI}(t)=[u_k^s(t)]^{IJ}[a_\mathbf{k}^s]^J+se^{-i\theta_\mathbf{k}}\cdot[v_k^s(t)^*]^{IJ}[a_{-\mathbf{k}}^{s\dagger}]^J
\end{equation}
where
\begin{eqnarray}
 \left[u_k^s(t)\right]^{IJ}
  &=& \delta^{IJ}\cdot\sqrt{\frac{1}{2}\left(1-\frac{s|\mathbf{k}|}{[\omega_k(t)]^{II}}\right)}e^{-i\int_{t_0}^tdt'[\omega_k(t')]^{II}} \\
 \left[v_k^s(t)\right]^{IJ}
  &=& \delta^{IJ}\cdot\sqrt{\frac{1}{2}\left(1+\frac{s|\mathbf{k}|}{[\omega_k(t)]^{II}}\right)}e^{-i\int_{t_0}^tdt'[\omega_k(t')]^{II}}
\end{eqnarray}
and $a_\mathbf{k}^s$ $(a_\mathbf{k}^{s\dagger})$ is an annihilation (a creation) operator which satisfies
\begin{equation}
 [[a_\mathbf{k}^s]^I, [a_{\mathbf{k}'}^{r\dagger}]^J] = (2\pi)^3\delta^3(\mathbf{k}-\mathbf{k}')\delta^{sr}\delta^{IJ},
 \qquad [[a_\mathbf{k}^s]^I, [a_{\mathbf{k}'}^r]^J] = 0.
\end{equation}
}
\begin{eqnarray}
 \langle\nu_\mathbf{k}^{sI\dagger}\nu_\mathbf{k}^{sJ}\rangle_{t=t_0}
  &=& V\cdot \frac{1}{2}\left(1+\frac{s|\mathbf{k}|}{[\omega_k(t_0)]^{II}}\right)\cdot\delta^{IJ} \\
 \qquad \langle\nu_\mathbf{-k}^{sI}\nu_\mathbf{-k}^{sJ\dagger}\rangle_{t=t_0}
  &=& V\cdot \frac{1}{2}\left(1-\frac{s|\mathbf{k}|}{[\omega_k(t_0)]^{II}}\right)\cdot\delta^{IJ} \\
 \qquad \langle\nu_\mathbf{-k}^{sI}\nu_\mathbf{k}^{sJ}\rangle_{t=t_0}
  &=& V\cdot \frac{[m_\nu(t_0)]^{II}}{2[\omega_k(t_0)]^{II}}\cdot\delta^{IJ}
\end{eqnarray}
where $t=t_0$ is an initial time, $V \equiv \int d^3x$ is a spatial volume of the system, and
\begin{eqnarray}
 [\omega_k(t)]^{II}
  &\equiv& \sqrt{|\mathbf{k}|^2+[m_\nu(t)^2]^{II}} \nonumber \\
  &=& \sqrt{|\mathbf{k}|^2+\frac{\langle h(t)\rangle^4}{\langle h_{\rm now}\rangle^4}([m_{\nu,{\rm now}}]^{II})^2}. \label{eq:energy_of_lhnu}
\end{eqnarray}

\subsection{Evolution of Higgs background} \label{sec:EoHB}
The time dependence of the matrices $m_\nu(t)$ and $Z(t)$ is described by the dynamics of the Higgs background.  Thus, we must follow the time evolution of the Higgs background in order to solve Eqs.(\ref{eq:eom_nu_bar_nu})-(\ref{eq:eom_nu_nu}).  The Lagrangian relating to the Higgs sector is given by
\begin{eqnarray}
 \mathcal{L}_{\rm Higgs}
  &=& D^\mu H^{a\dagger}D_\mu H^a - \frac{1}{4}\lambda(H^{a\dagger}H^a)^2 + \mathcal{L}_{\rm Yukawa}
\end{eqnarray}
where $\mathcal{L}_{\rm Yukawa}$ is the Higgs interaction terms with fermions.
We neglect the quadratic term of the Higgs because the scale we focus is much higher than the electro-weak scale.
Using the unitary gauge, $H^1=0, \: H^2=h/\sqrt{2}$, the Lagrangian in each component can be represented as
\begin{eqnarray}
 \mathcal{L}_{\rm Higgs}
  &=& \frac{1}{2}\partial^\mu h\partial_\mu h + \frac{1}{8}(g_Y^2+g_W^2)Z^\mu Z_\mu h^2 \nonumber\\
  & & +\frac{1}{4}g_W^2W^{+\mu}W_\mu^-h^2 -\frac{1}{4}\lambda h^4 + \mathcal{L}_{\rm Yukawa} \label{eq:Lagrangian_higgs}
\end{eqnarray}
where we denote
\begin{equation}
 \left(\begin{array}{c}Z_\mu\\A_\mu\end{array}\right)
  \equiv\left(\begin{array}{cc}\cos\theta&-\sin\theta\\\sin\theta&\cos\theta\end{array}\right)
  \left(\begin{array}{c}W_\mu^3\\B_\mu\end{array}\right),
\end{equation}
\begin{equation}
 \sin\theta\equiv\frac{g_Y}{\sqrt{g_Y^2+g_W^2}},\quad\cos\theta\equiv\frac{g_W}{\sqrt{g_Y^2+g_W^2}},
\end{equation}
\begin{equation}
 W_\mu^{\pm}\equiv\frac{1}{\sqrt{2}}(W_\mu^1\mp iW_\mu^2).
\end{equation}
The Lagrangian (\ref{eq:Lagrangian_higgs}) leads the equation of motion for the Higgs as
\begin{eqnarray}
 0 &=& \partial_t^2 h -\partial_i\partial_i h + \lambda h^3 \nonumber \\
  & &- \left(\frac{1}{4}(g_Y^2+g_W^2)Z^\mu Z_\mu
   + \frac{1}{2}g_W^2W^{+\mu}W_\mu^-\right)h \nonumber \\
  & & +({\rm fermion\mathchar`'s \:\: terms}).
\end{eqnarray}
Taking the vacuum expectation value in the above equation, we can obtain the equation for the Higgs background as
\begin{eqnarray}
 0 &=& \partial_t^2 \langle h\rangle +\lambda \langle h\rangle^3+J_{\rm BR} \label{eq:eom_h_bkg1}
\end{eqnarray}
where $J_{\rm BR}$ is a backreaction term that consists of
\begin{eqnarray}
 J_{\rm BR}
  &\equiv& \delta m_h^2 \cdot \langle h \rangle 
    +({\rm fermion\mathchar`'s \:\: two\mathchar`-point \:\: functions}) \nonumber \\
   & & +\lambda\langle\tilde{h}^3\rangle-\frac{1}{4}(g_Y^2+g_W^2)\langle Z^\mu Z_\mu \tilde{h}\rangle \nonumber \\
   & & - \frac{1}{2}g_W^2\langle W^{+\mu}W_\mu^-\tilde{h}\rangle, \label{eq:backreaction_term}\\
 \tilde{h}
  &\equiv& h-\langle h\rangle, \\
 \delta m_h^2
  &\equiv& 3\lambda\langle\tilde{h}^2\rangle
   -\frac{1}{4}(g_Y^2+g_W^2)\langle Z^\mu Z_\mu \rangle - \frac{1}{2}g_W^2\langle W^{+\mu}W_\mu^-\rangle. \nonumber \\ \label{eq:delta_mh}
\end{eqnarray}
This backreaction plays a quite important role.  As we will see in section \ref{sec:numerical_result_case_A}, the lepton asymmetry cannot be fixed without the backreaction.

Because the backreaction (\ref{eq:backreaction_term}) includes too much information and thus the form is complicated, let us extract the relevant effect and approximate them to be a useful form.  At first, we neglect three-point functions since we constructed the differential equations up to two-point functions.  This approximation is valid as long as $\langle Z^\mu\rangle\sim\langle W^{\pm\mu}\rangle\sim0$.
Moreover, we can expect that the bosonic two-point functions would be much more significant than the fermionic functions
because each two-point function relates to the number density and the bosons would be exponentially produced due to the parametric resonance by the coherent oscillation of the Higgs background.
Therefore, we can approximate the backreaction term as
\begin{equation}
 J_{\rm BR} \sim \delta m_h^2 \cdot \langle h \rangle.
\end{equation}
Speaking roughly, $\delta m_h^2$ is the product of the couplings and the total number densities of the Higgs and the gauge bosons.  Because it is complicated to follow the time evolution of the two-point functions of these species (\ref{eq:delta_mh}), we approximate them by a single real scalar field\footnote{The scalar $\chi$ defined here is NOT a new field in the type-I seesaw model.  We treat this scalar as an approximation technique to estimate the degrees of freedom generated by the oscillating Higgs background.
} as
\begin{equation}
 \delta m_h^2 \sim \mathcal{N}_{\rm deg}\cdot\frac{1}{4}g_W^2 \langle \chi^2 \rangle \label{eq:approximation_delta_m_h^2}
\end{equation}
where $\chi$ is an artificial scalar field that satisfies
\begin{equation}
 0 = \partial_t^2\chi-\partial_i\partial_i\chi+\frac{1}{4}g_W^2\langle h\rangle^2 \chi, \qquad \langle \chi \rangle =0,
\end{equation}
and
\begin{eqnarray}
 \mathcal{N}_{\rm deg}
  &\equiv& \frac{12\lambda}{g_W^2}\frac{\langle\tilde{h}^2\rangle}{\langle \chi^2\rangle} -\frac{g_Y^2+g_W^2}{g_W^2}\frac{\langle Z^\mu Z_\mu \rangle}{\langle \chi^2\rangle} - \frac{2\langle W^{+\mu}W_\mu^-\rangle}{\langle \chi^2\rangle} \nonumber \\\\
  &\sim& \frac{12\lambda}{g_W^2}\cdot 1+\left(1+\frac{g_Y^2}{g_W^2}\right)\cdot 3+6 \label{eq:effective_degrees}
\end{eqnarray}
is an effective degrees of freedom included in (\ref{eq:delta_mh}).
The coefficients $1, 3, 6$ in (\ref{eq:effective_degrees}) correspond to the degrees of freedom of $h, Z, W^{\pm}$ bosons, respectively.

Finally, the differential equation for the Higgs background can be derived as\footnote{
The representation
\begin{equation}
 \chi(t,\vec{x})=\int\frac{d^3k}{(2\pi)^3}e^{i\vec{k}\cdot\vec{x}}\left(u_k(t)a_{\vec{k}}+u_k(t)^*a_{-\vec{k}}^\dagger\right)
\end{equation}
leads to
\begin{equation}
 \langle \chi^2\rangle=\int\frac{d^3k}{(2\pi)^3a^3}|u_k(t)|^2,
\end{equation}
but this term diverges.
In order to renormalize, we add a counter term as
\begin{equation}
 \langle \chi^2\rangle \:\: \rightarrow \:\: \langle\chi^2\rangle_{\rm ren}=\int\frac{d^3k}{(2\pi)^3a^3}\left(|u_k(t)|^2-\frac{1}{2\omega_{\chi k}(t)}\right)
\end{equation}
by hand.  The counter term $1/2\omega_{\chi k}$ corresponds to $\langle 0(t)|\chi^2|0(t)\rangle$
where $|0(t)\rangle$ is a vacuum state defined at time $t$.
}
\begin{eqnarray}
 0 &=& \partial_t^2 \langle h\rangle + \lambda \langle h\rangle^3 +\mathcal{N}_{\rm deg}\cdot\frac{1}{4}g_W^2\langle \chi^2\rangle\langle h\rangle \\
   &=& \partial_t^2 \langle h\rangle + \lambda \langle h\rangle^3 \nonumber \\
   & & +\mathcal{N}_{\rm deg}\cdot\frac{1}{4}g_W^2\int\frac{d^3k}{(2\pi)^3}\left(|u_k|^2-\frac{1}{2\omega_{\chi k}}\right)\langle h\rangle \label{eq:eom_higgs_bkg}
\end{eqnarray}
where $u_k$ is $\chi$'s time-dependent wave function which satisfies
\begin{equation}
 0 = \partial_t^2{u}_k+\omega_{\chi k}^2u_k, \quad \omega_{\chi k}\equiv\sqrt{|\mathbf{k}|^2+\frac{1}{4}g_W^2\langle h\rangle^2}, \label{eq:eom_u_k}
\end{equation}
\begin{equation}
 u_k(t_0) = \frac{1}{\sqrt{2\omega_{\chi k}(t_0)}}, \quad \dot{u}_k(t_0) =-i\omega_k(t_0)u_k(t_0).
\end{equation}
The above initial conditions for $u_k$ indicate zero-particle state as the initial state.  Thus, this analysis is valid in the case that the thermal particle number described by the temperature of the Universe is negligible.  We assume that the produced bosons due to the oscillating Higgs background are more than the thermal particles.
Otherwise, the backreaction does not sufficiently affect to the Higgs background and the final lepton number would not be fixed.

\subsection{Scales of the particle production}
In this section, we discuss what the momentum scale of the produced left-handed neutrinos is .
Non-perturbative particle production occurs when the adiabatic condition is violated.
The condition for producing the heaviest left-handed neutrino is given by
\begin{equation}
 1 \lesssim \left|\frac{[\dot{\omega}_k]_{\rm heaviest}}{([\omega_k]_{\rm heaviest})^2}\right|_{\mathbf{k}\sim0}
  \sim \frac{2\langle h_{\rm now}\rangle^2}{m_{\nu,{\rm heaviest}}}
   \cdot\left|\frac{\langle \dot{h}\rangle}{\langle h\rangle^3}\right| \label{eq:non-ad_cond1}
\end{equation}
where $[\omega_k]_{\rm heaviest}\equiv[\omega_k]^{II}$ and $m_{\nu,{\rm heaviest}}\equiv[m_{\nu,{\rm now}}]^{II}$ for the heaviest generation $I$ of the left-handed neutrinos.  In order to obtain the production scale, we need to know the dynamics of the Higgs background.  At the beginning of the particle production era, the backreaction $J_{\rm BR}$ in (\ref{eq:eom_h_bkg1}) can be neglected.  Then the time derivative of the Higgs background can be represented as
\begin{equation}
 |\langle \dot{h}(t)\rangle|\sim\sqrt{\frac{\lambda}{2}\left(\langle h_{\rm max}\rangle^4-\langle h(t)\rangle^4\right)}. \label{eq:dot_h_relation}
\end{equation}
In the derivation of the above equation, we assume $\langle \dot{h}\rangle=0$ when $\langle h\rangle=\langle h_{\rm max}\rangle$.
Substituting (\ref{eq:dot_h_relation}) into (\ref{eq:non-ad_cond1}), we obtain
\begin{eqnarray}
 1 \lesssim \frac{2^{3/2}}{\sqrt{Q}} \cdot\frac{\sqrt{1-(\langle h(t)\rangle / \langle h_{\rm max}\rangle)^4}}{|\langle h(t)\rangle / \langle h_{\rm max}\rangle|^3} \label{eq:non_adiabatic_condition}
\end{eqnarray}
where
\begin{eqnarray}
 Q &\equiv& \frac{4}{\lambda}\left(\frac{m_{\nu,{\rm heviest}}\langle h_{\rm max}\rangle}{\langle h_{\rm now}\rangle^2}\right)^2 \nonumber \\
  &=& \frac{4}{\lambda}\cdot\left(\frac{m_{\nu,{\rm heviest}}}{0.1\:{\rm eV}}\frac{\langle h_{\rm max}\rangle}{6.05\times 10^{14}\:{\rm GeV}}\right)^2. \label{eq:q_parameter}
\end{eqnarray}
Note that the parameter $Q$ corresponds to the resonance parameter $q$ known in the Mathieu equation\footnote{The Mathieu equation for a function $y=y(x)$ is given by
\begin{equation}
 0=y''+(A-2q\cos 2x)y
\end{equation}
where $A$ and $q$ are the resonance parameters.  In this equation, the non-adiabatic condition is obtained as
\begin{equation}
 1\lesssim \left|\frac{(\sqrt{A-2q\cos 2x})'}{A-2q\cos 2x}\right|_{A=2q}=\frac{1}{2\sqrt{q}}\cdot\frac{|\cos x|}{\sin^2x}.
\end{equation}
}.

In the case of $Q\gg1$ which corresponds to the broad resonance, the condition (\ref{eq:non_adiabatic_condition}) can be simplified to
\begin{equation}
 |\langle h(t)\rangle|\lesssim \sqrt{2} Q^{-1/6} \langle h_{\rm max}\rangle.
\end{equation}
This result shows us the production area of the left-handed neutrinos.  Furthermore, applying the Tayler expansion to the Higgs background
\begin{equation}
 \langle h(t)\rangle\sim\langle \dot{h}(t=t_*)\rangle(t-t_*)\sim\sqrt{\frac{\lambda}{2}}\langle h_{\rm max}\rangle^2(t-t_*)
\end{equation}
where $t=t_*$ is a time when $\langle h(t=t_*)\rangle=0$,
we obtain the time scale of the particle production around $\langle h\rangle=0$ as
\begin{equation}
 |t-t_*| \lesssim \left(Q^{1/6}\sqrt{\frac{\lambda}{4}}\langle h_{\rm max}\rangle\right)^{-1}\equiv \Delta t. \label{eq:PP_time_scale}
\end{equation}
This time scale implies the momentum scale of the produced particles as
\begin{equation}
 |\mathbf{k}|\lesssim |[m_\nu(t=t_*+\Delta t)]_{\rm heaviest}|=\frac{1}{\Delta t}\equiv \Delta k.
\end{equation}

In the case of $Q\lesssim 1$ which corresponds to the narrow resonance, the condition (\ref{eq:non_adiabatic_condition}) is satisfied for almost any value of $\langle h\rangle$ except the narrow area of $\langle h\rangle \sim \langle h_{\rm max}\rangle$.   Therefore, the time scale of the particle production is estimated by the oscillation time scale of the Higgs background $t_{\rm osc}$ as
\begin{eqnarray}
 \Delta t \sim t_{\rm osc}
  &\sim& \int_0^{\langle h_{\rm max}\rangle}dh\frac{1}{\sqrt{\frac{\lambda}{2}\left(\langle h_{\rm max}\rangle^4-h^4\right)}} \nonumber \\
  &=& 0.9270\cdot\left(\sqrt{\frac{\lambda}{4}}\langle h_{\rm max}\rangle\right)^{-1} \label{eq:osc_time_scale}
\end{eqnarray}
and thus the momentum scale can be estimated by
\begin{equation}
 |\mathbf{k}|\lesssim |[m_\nu(\langle h\rangle=\langle h_{\rm max}\rangle)]_{\rm heaviest}|=\frac{m_{\nu,{\rm heviest}}\langle h_{\rm max}\rangle^2}{\langle h_{\rm now}\rangle^2}.
\end{equation}

\subsection{Entropy density}
Since the cosmological observation of the baryon number density $n_B$ is normalized by the entropy density $s$ as $n_B/s$, we need to estimate not only the net lepton number density but also the entropy density.
In this section, we discuss the produced entropy in cases of the following two types of situation:
\begin{itemize}
 \item Case A: The main contribution of the entropy is produced by the perturbative decay of the inflation field $\phi$.
 \item Case B: The main contribution of the entropy is produced by the non-perturbative decay of the Higgs background due to its oscillation, that is, the lepton number generation and the entropy production occurs simultaneously.
\end{itemize}
The difference in Cases A and B are not only the dominant component of the Universe in the epoch of entropy production but also the entropy production mechanism. In many cases, the situation is included in Case A.  While in Case B, the Universe must be dominant by the Higgs background.  To realize this situation seems difficult if the model includes both the Higgs background and any inflation fields.  As an example of Case B, the Higgs inflation model \cite{Bezrukov:2007ep} would be applicable.

\subsubsection{Case A: entropy production by inflation field} \label{sec:case_A}
The produced entropy density $s$ at the decay time of the inflation field $t=t_R$ is given by
\begin{equation}
 s(t_R) = \left(\frac{g_*}{180\pi}\right)^{1/4}\left(\Gamma_\phi M_{pl}\right)^{3/2}=0.65 \cdot \left(\Gamma_\phi M_{pl}\right)^{3/2}
\end{equation}
where $g_*\sim 100$ is the degrees of freedom of the relativistic particles, $\Gamma_\phi=1/t_R$ is a decay width of $\phi$ and $M_{pl}=1.22\times 10^{19}$ GeV is the Planck mass.  Assuming the lepton number generation is completed at $t=t_L<t_R$, the lepton number density $n_L$ at $t=t_R$ can be written as
\begin{eqnarray}
 n_L(t_R)
  &=& n_L(t_L) \cdot \frac{a(t_L)^3}{a(t_R)^3} = n_L(t_L) \cdot \Gamma_\phi^2 t_L^2
\end{eqnarray}
because of $n_L(t)\propto 1/a(t)^3\propto t^{-2}$ until $t=t_R$.  Therefore, the lepton-to-entropy ratio at $t=t_R$ can be shown as
\begin{eqnarray}
 \frac{n_L(t_R)}{s(t_R)}
 &=& \frac{n_L(t_L)\cdot \Gamma_\phi^2t_L^2}{0.65\cdot(\Gamma_\phi M_{pl})^{3/2}} \\
 &=& 4.0\times 10^{-10}\cdot \frac{n_L(t_L)\Delta t^3}{10^{-8}} \left(\frac{\langle h_{\rm max}
\rangle}{10^{16} \ {\rm GeV}}\right)^2 \nonumber \\
  & & \qquad \qquad \quad \cdot \left(\frac{\Gamma_\phi}{10^{12} \ {\rm GeV}}\right)^{1/2}\left(\frac{N_{\rm osc}}{5}\right)^2 \label{eq:lepton_to_entropy_case_A}
\end{eqnarray}
where $\Delta t$ is a production time scale of the left-handed neutrino defined in (\ref{eq:PP_time_scale}) and
\begin{equation}
 N_{\rm osc}\sim\frac{1}{2\pi}\cdot\frac{t_L}{4t_{\rm osc}}
\end{equation}
is the number of the Higgs oscillation until the lepton number generation completes.  Hence, this scenario requires a high scale amplitude around $\langle h_{\rm max}\rangle\sim 10^{16}$ GeV and strong decay of the inflation field.  We will see later with numerical results whether enough lepton numbers are produced or not.

\subsubsection{Case B: entropy production by the Higgs background} \label{sec:case_B}
In this situation, we need to evaluate the entropy from the information of the decay products through the non-perturbative decay from the Higgs oscillation.
In our analysis, we estimate the entropy density with the distribution functions of the produced particles as\footnote{In the case of the equilibrium distribution
\begin{equation}
 f_k = \frac{1}{e^{(\omega_k-\mu)/T}\mp 1} \qquad (-:{\rm bosons}, \:+:{\rm fermions})
\end{equation}
where $\omega_k=\sqrt{|\mathbf{k}|^2+m^2}$ is one-particle energy, $\mu$ is chemical potential  and $T$ is temperature, one can derive the familiar representation of the entropy density from (\ref{eq:entropy_def}) as
\begin{equation}
 s=\frac{\rho+p-\mu n}{T}
\end{equation}
where $\rho$ is the energy density, $p$ is the pressure and $n$ is the number density.}
\begin{eqnarray}
 s &=& \sum_{i\:\:{\rm for\:\:bosons}}\mathcal{N}_i\int\frac{d^3k}{(2\pi)^3}\left[(1+f_k^{(i)})\ln(1+f_k^{(i)})\right.\nonumber \\
  & & \qquad \qquad \qquad \qquad \qquad \qquad \qquad \left. -f_k^{(i)}\ln f_k^{(i)}\right] \nonumber \\
   & & + \sum_{i\:\:{\rm for\:\:fermions}}\mathcal{N}_i\int\frac{d^3k}{(2\pi)^3}\left[-(1-f_k^{(i)})\ln(1-f_k^{(i)})\right.\nonumber \\
  & & \qquad \qquad \qquad \qquad \qquad \qquad \qquad \left. -f_k^{(i)}\ln f_k^{(i)}\right] \label{eq:entropy_def}
\end{eqnarray}
where $\mathcal{N}_i$ and $f_k^{(i)}$ are the degrees of freedom and the distribution function for species $i$ particle, respectively.

In our scenario, sizable gauge bosons $W^\pm$ and $Z$ and the Higgs bosons are produced by the oscillating Higgs background.  
As we approximated in (\ref{eq:approximation_delta_m_h^2}), we can also approximate (\ref{eq:entropy_def}) by the effective degrees of freedom $\mathcal{N}_{\rm deg}$ and the distribution of $\chi$ particle.  Then, we can obtain
\begin{equation}
 s \sim \mathcal{N}_{\rm deg}\int\frac{d^3k}{(2\pi)^3}\left[(1+f_k^{(\chi)})\ln(1+f_k^{(\chi)})-f_k^{(\chi)}\ln f_k^{(\chi)}\right] \label{eq:entropy_rep}
\end{equation}
where the distribution function can be represented by the wave function as
\begin{equation}
 f_k^{(\chi)} = \frac{|\dot{u}_k|^2+\omega_{\chi k}^2|u_k|^2}{2\omega_{\chi k}}-\frac{1}{2}. \label{eq:distribution_rep}
\end{equation}
Since it is difficult to obtain the analytic results in this case, we will see the numerical results in section \ref{sec:NR}.

\subsection{Model parameters}
Before we show our numerical results, we mention the required input parameters for our analysis.  
There are 17 model parameters required by the analysis of the lepton asymmetry:
\begin{itemize}
 \item The gauge couplings $g_Y, g_W$ and the Higgs self-coupling $\lambda$
 \item Masses of the left-handed neutrinos $m_{\nu,{\rm now}}={\rm diag}(m_1,m_2,m_3)$
 \item Masses of the right-handed neutrinos $M_R={\rm diag}(M_1,M_2,M_3)$
 \item Complex orthogonal matrix $O$: 6 real parameters
 \item Initial values of $\langle h(t_0)\rangle, \langle \dot{h}(t_0)\rangle$
\end{itemize}
The SM parameters $g_Y, g_W$, and $\lambda$ could be determined by the renormalization group running once the initial conditions for $\langle h(t_0)\rangle, \langle \dot{h}(t_0)\rangle$ are determined. 

For simplicity, we assume that $m_3$ is the heaviest left-handed neutrino and that a non-degenerate mass spectrum to the left-handed neutrinos in later analysis\footnote{Although the case of the degenerate mass spectrum is also applicable to our scenario, we do not consider such a case in this paper because the mass scale of the left-handed neutrinos cannot be determined.}. 
Taking into account the observations of the neutrino oscillation\cite{Abe:2019vii}, we set the heaviest mass of the left-handed neutrinos as
\begin{equation}
 m_3\sim\sqrt{|\Delta m_{32}^2|}=\sqrt{2.5\times 10^{-3}} \:\:{\rm eV}.
\end{equation}

The mass scale of the right-handed neutrinos is constrained in our analysis.  The lightest right-handed neutrino must be much heavier than the heaviest left-handed neutrino because of the validity of the effective theory.  Hence, the lightest mass of the right-handed neutrinos must be
\begin{eqnarray}
 [M_R]_{\rm lightest}
  &\gg& {\rm max}([m_\nu(t)]_{\rm heviest}) \\
  & & = \frac{\langle h_{\rm max}\rangle^2}{\langle h_{\rm now}\rangle^2}m_3=\frac{\langle h_{\rm max}\rangle^2}{1.21\times 10^{15}\:{\rm GeV}} \label{eq:bound_rhnm}
\end{eqnarray}
where $\langle h_{\rm max}\rangle$ is a maximal value of $|\langle h(t)\rangle|$.
Note that the typical scale in our scenario is characterized by
\begin{equation}
 \frac{\langle h_{\rm now}\rangle^2}{m_3}=1.21\times 10^{15}\:{\rm GeV}.
\end{equation}

Finally, we mention the treatment of the complex orthogonal matrix $O$.  Since this matrix does not have any constraints, we treat it as a set of free parameters.  The parametrization can be taken as
\begin{eqnarray}
 O &=& \left(\begin{array}{ccc} 1 & 0 & 0 \\ 0 & c_{23} & s_{23} \\ 0 & -s_{23} & c_{23} \end{array}\right)
  \left(\begin{array}{ccc} c_{13} & 0 & -s_{13} \\ 0 & 1 & 0 \\ s_{13} & 0 & c_{13} \end{array}\right)
  \left(\begin{array}{ccc} c_{12} & s_{12} & 0 \\ -s_{12} & c_{12} & 0 \\ 0 & 0 & 1 \end{array}\right) \nonumber \\ \label{eq:parametrization_of_o}
\end{eqnarray}
where
\begin{eqnarray}
 c_{ij} &\equiv& \cos\theta_{ij} \nonumber \\
  &=& \cosh({\rm Im}\:\theta_{ij})\cdot \cos({\rm Re}\:\theta_{ij})-i\sinh({\rm Im}\:\theta_{ij})\cdot \sin({\rm Re}\:\theta_{ij}) \nonumber \\\\
 s_{ij} &\equiv& \sin\theta_{ij} \nonumber \\
  &=& \cosh({\rm Im}\:\theta_{ij})\cdot \sin({\rm Re}\:\theta_{ij})+i\sinh({\rm Im}\:\theta_{ij})\cdot \cos({\rm Re}\:\theta_{ij}). \nonumber \\ \label{eq:parametrization_of_sine}
\end{eqnarray}
The complex parameters $\theta_{12}, \theta_{23}, \theta_{13}$ correspond to 6 real parameters.

\subsection{Numerical results} \label{sec:NR}
Finally, we show our numerical results with a set of specific parameters.
In our analysis, we set the gauge couplings and the Higgs self-coupling at the Higgs oscillation scale as\footnote{For the Higgs self-coupling $\lambda$, there exists the vacuum stability problem that the coupling $\lambda$ tends to run into negative at high scale $\gtrsim 10^{9-15}$ GeV\cite{Degrassi:2012ry,Buttazzo:2013uya}.  Since our scenario requires the oscillation of the Higgs background, we assume that the Higgs self-coupling maintains positive at the focusing scale.
}
\begin{equation}
 g_Y^2 = g_W^2 = 4\pi\cdot\frac{1}{40}, \qquad \lambda = 0.001. \label{eq:value_couplings}
\end{equation}
Then, the effective degrees of freedom defined in (\ref{eq:effective_degrees}) is evaluated as
\begin{equation}
 \mathcal{N}_{\rm deg} \sim 12.
\end{equation}
Basically, the other parameters are free.  In this paper, we choose the parameters to be
\begin{eqnarray}
 m_{\nu,{\rm now}} &=& {\rm diag} (0, \sqrt{\Delta m_{21}^2}, \sqrt{|\Delta m_{32}^2|}) \nonumber \\
   &=& {\rm diag} (0, \sqrt{7.5\times 10^{-5}}, \sqrt{2.5\times 10^{-3}}) \:\: {\rm eV}, \\
 M_R &=& M_1 \times {\rm diag} (1, 10, 100),
\end{eqnarray}
\begin{equation}
 \theta_{12} = \frac{\pi}{6}+0.1i, \quad
 \theta_{23} = \frac{\pi}{12}+0.2i, \quad
 \theta_{13} = \frac{\pi}{4}+0.3i,
\end{equation}
where $m_{\nu,{\rm now}}$ and $M_R$ are diagonal mass matrices of the left- and right-handed neutrinos on $\langle h\rangle=246$ GeV and each $\theta_{ij}$ defined in (\ref{eq:parametrization_of_o})-(\ref{eq:parametrization_of_sine}) is a parameter of the orthogonal matrix $O$.  Furthermore, we assume
\begin{equation}
 \langle \dot{h}(t_0)\rangle=0
\end{equation}
for simplicity.  With this assumption, we can regard $\langle h_{\rm max}\rangle$ as $\langle h(t_0)\rangle$.  The rest parameters $\langle h(t_0)\rangle$ and $M_1$ are treated as variables in our analysis.  Using the above parameters, we solve (\ref{eq:eom_nu_bar_nu})-(\ref{eq:eom_nu_nu}) and (\ref{eq:eom_higgs_bkg}) numerically.  Substituting the obtained values into (\ref{eq:lepton_num_rep}), (\ref{eq:entropy_rep}) and (\ref{eq:distribution_rep}) at each time, we can follow the time evolution of the net lepton number.

\subsubsection{Case A: entropy production by inflation field} \label{sec:numerical_result_case_A}
\begin{figure}[t]
 \begin{center}
  \includegraphics[scale=0.37]{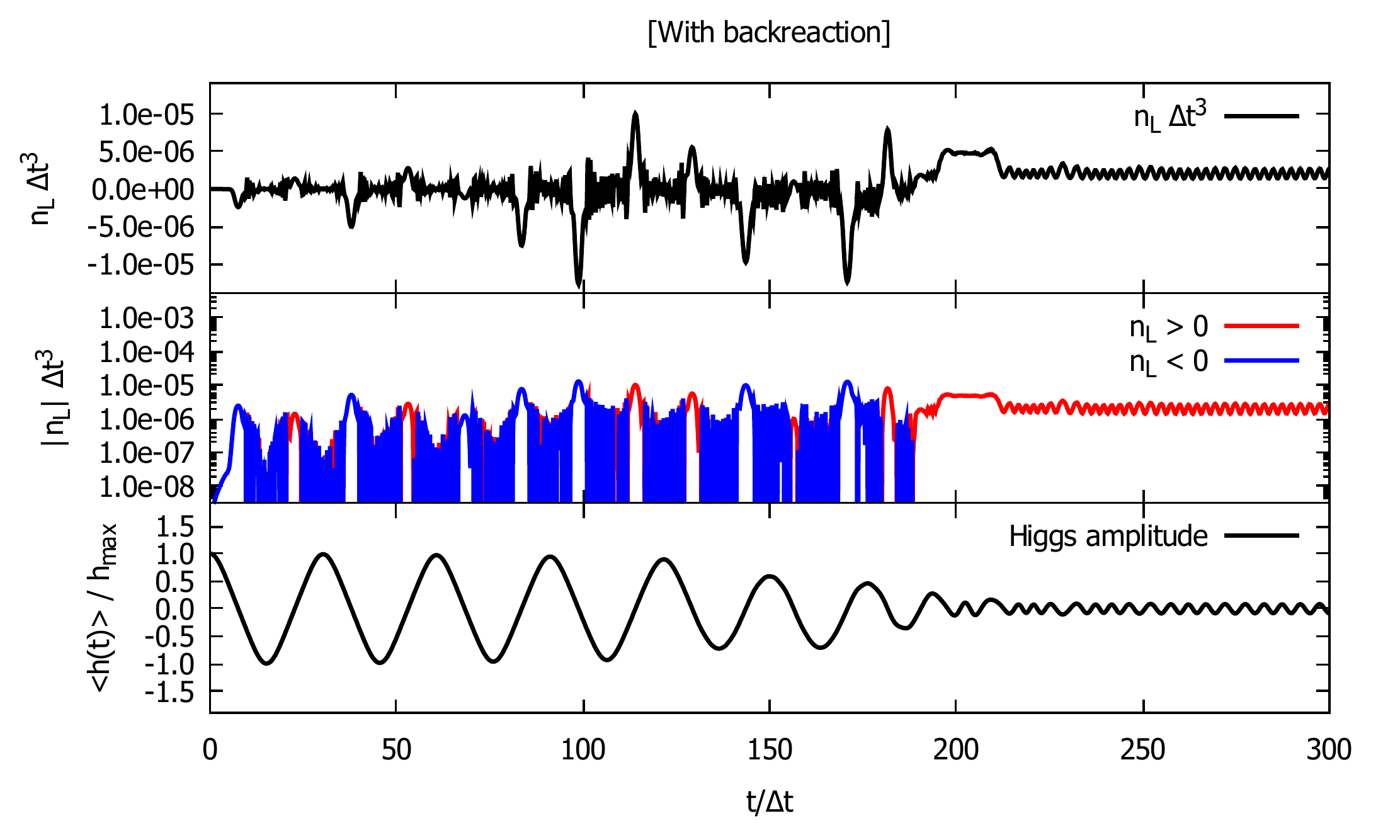}
  \includegraphics[scale=0.37]{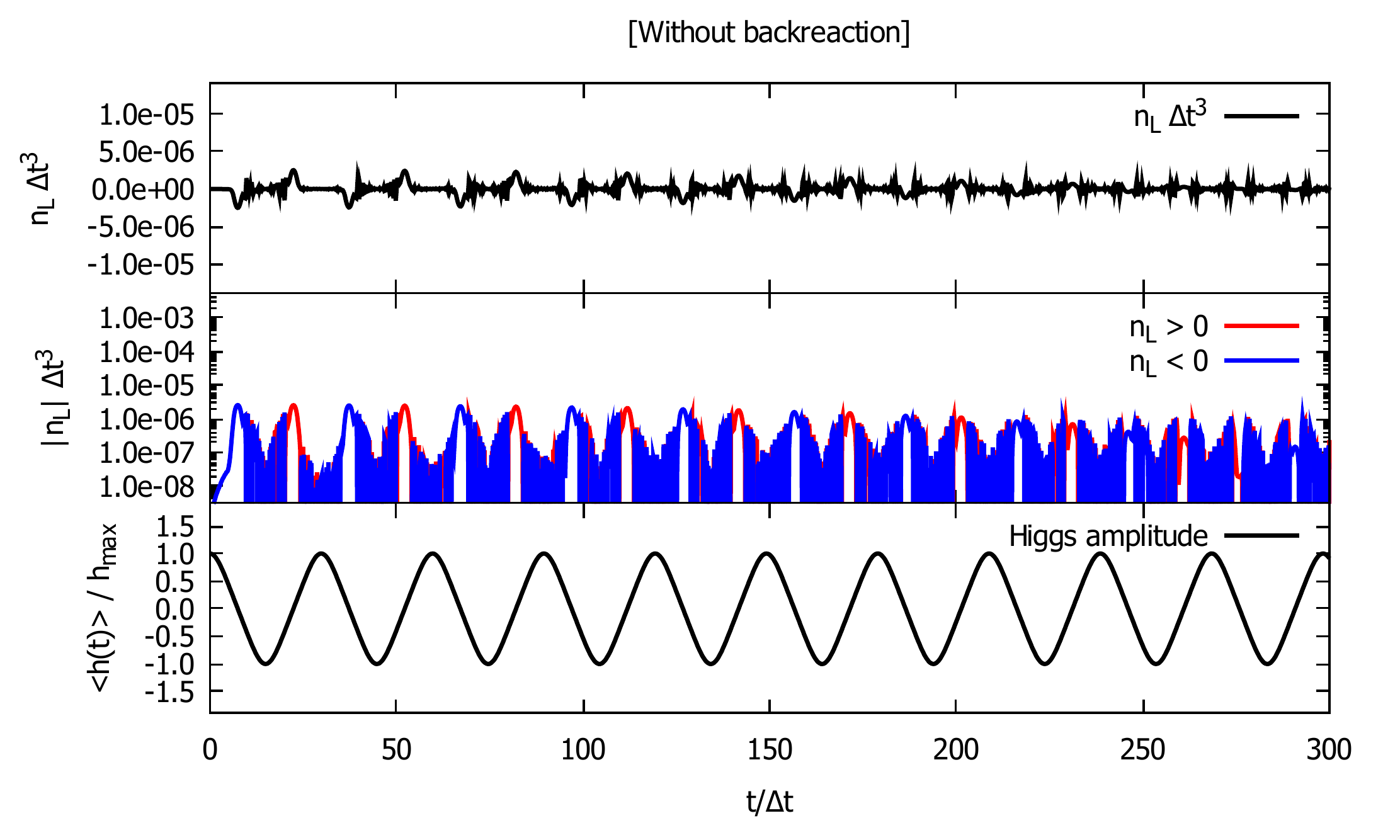}
  \caption{{\bf Upper:} The time evolutions of the lepton number in a volume of $\Delta t^3$ and the amplitude of the Higgs background.  The parameters are set by $\langle h(t_0)\rangle=10^{16}$ GeV and $M_1=10^{17}$ GeV.  The first two figure shows the time evolution of the net lepton number with normal and logarithmic scales of the vertical axes.  The third figure shows the time evolution of the Higgs amplitude.  {\bf Lower:} The time evolutions with the same parameters to the upper panel but without the backreaction effect ($J_{\rm BR}=0$ in (\ref{eq:eom_h_bkg1})).}
  \label{fig:result_higher_amplitude}
 \end{center}
\end{figure}

At first, we focus on a situation of Case A in which the entropy production is induced by the decay of the inflation field after the generation of the lepton asymmetry.  As discussed in section \ref{sec:case_A}, $h_{\rm max}\sim 10^{16}$ GeV is required.  Our numerical results with
\begin{equation}
 \langle h(t_0)\rangle = 10^{16}\:\:{\rm GeV}, \quad  M_1 = 10^{17} \:\:{\rm GeV} \label{eq:parameters_higher}
\end{equation}
are shown in the upper panel of Figure \ref{fig:result_higher_amplitude}.  One can see that the evolution of the net lepton number oscillates between the positive and negative with a similar magnitude of $10^{-6}$ at the first stage.  The flipping of the sign happens when the Higgs amplitude reaches to the edge, i.e., $\langle h(t)\rangle \sim \langle h_{\rm max}\rangle$.  After the Higgs amplitude dumps at $t\sim 190\Delta t$ where $\Delta t$ is a time scale of the particle production defined in (\ref{eq:PP_time_scale}), however, the oscillation of the lepton number stops and its evolution freezes around
\begin{equation}
 n_L \Delta t^3 \sim 2\times 10^{-6}. \label{eq:numerical_result_net_lepton_higher_amp}
\end{equation}

The main reason for the amplitude reduction of the Higgs background is the resonant production of the gauge bosons and the Higgs boson.  The production of the left-handed neutrinos with lepton number violation also occurs but the energy conversion from the Higgs background is much smaller because the interaction is suppressed by the right-handed neutrino mass scale.  Once enough bosons are produced, their plasma behaves as an effective mass of the Higgs background. In consequence, the Higgs background loses its energy and the non-adiabatic condition, and hence the all of the particle production finally stops.  Since the neutrino production freezes, the asymmetry flipping also freezes.  As seen in the lower panel of Figure \ref{fig:result_higher_amplitude}, the asymmetry flipping lasts forever if the backreaction is not taken into account in the analysis.  Therefore, the effect of the backreaction has an important role to fix the final amount of the asymmetry.

Substituting the numerical results (\ref{eq:numerical_result_net_lepton_higher_amp}) and $N_{\rm osc} \sim 7$ into (\ref{eq:lepton_to_entropy_case_A}), one can obtain
\begin{eqnarray}
 \frac{n_L(t_R)}{s(t_R)}
  &\sim& 2.4\times 10^{-10}\cdot\left(\frac{\Gamma_\phi}{4.2\times 10^{6} \ {\rm GeV}}\right)^{1/2}.
\end{eqnarray}
Thus, there exist possible parameters to explain in Case A if the amplitude of the Higgs background and the right-handed neutrino mass scale is higher.  Note that the estimation here is valid in the case that the effect of the decay product from the inflation field can be neglected during the generation of the lepton asymmetry.  Because the backreaction to the Higgs background in our numerical analysis only includes the contribution from $W^\pm$ and $Z$.  If the contribution of other decay products is not negligible, the time evolution of the Higgs background, and hence, the net lepton number would change.

In contrast to the case of parameters (\ref{eq:parameters_higher}), cases of the lower scales of $\langle h(t_0)\rangle$ and $M_1$ tend not to be able to explain the current asymmetry.  According to our calculation with parameters $\langle h(t_0)\rangle = 1.5\times 10^{14}$ GeV and $M_1 = 10^{15}$ GeV, the analysis shows $n_L(t_L)\Delta t^3 \sim -4\times 10^{-9}$ and $N_{\rm osc}\sim 5$.  (See the lower panel of Figure \ref{fig:comparison_asym} for its evolution.)  These results lead
\begin{eqnarray}
 \frac{n_L(t_R)}{s(t_R)}
  &\sim& 2.4\times 10^{-10}\cdot\left(\frac{\Gamma_\phi}{1.4\times 10^{15} \ {\rm GeV}}\right)^{1/2}. \label{eq:lepton_not_suitable}
\end{eqnarray}
Because the required decay width is greater than the mass of the inflation field $m_\phi\sim 10^{13}$ GeV, this scenario seems not to work.

\subsubsection{Case B: entropy production by the Higgs background}
\begin{figure}[t]
 \begin{center}
  \includegraphics[scale=0.37]{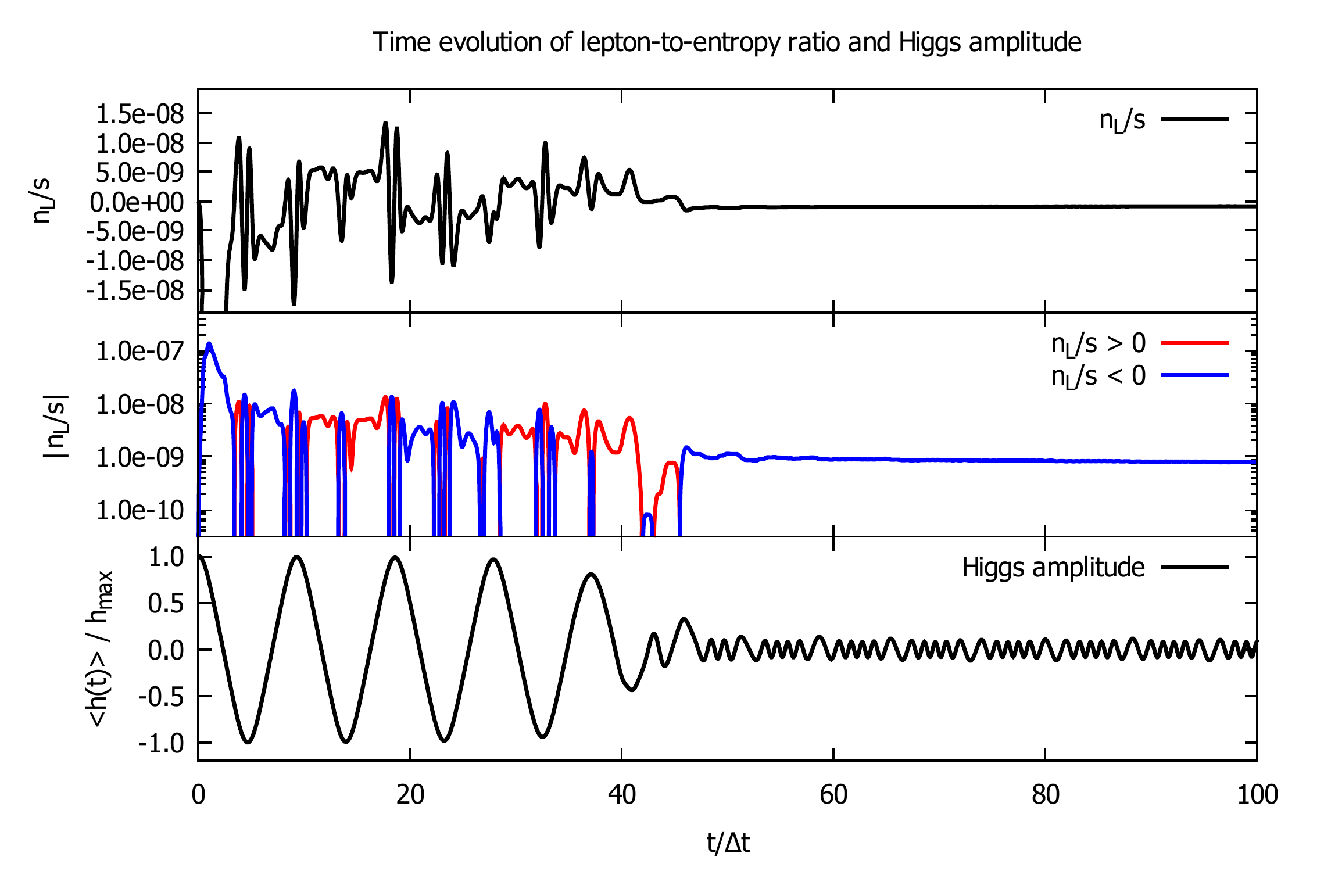}
  \includegraphics[scale=0.37]{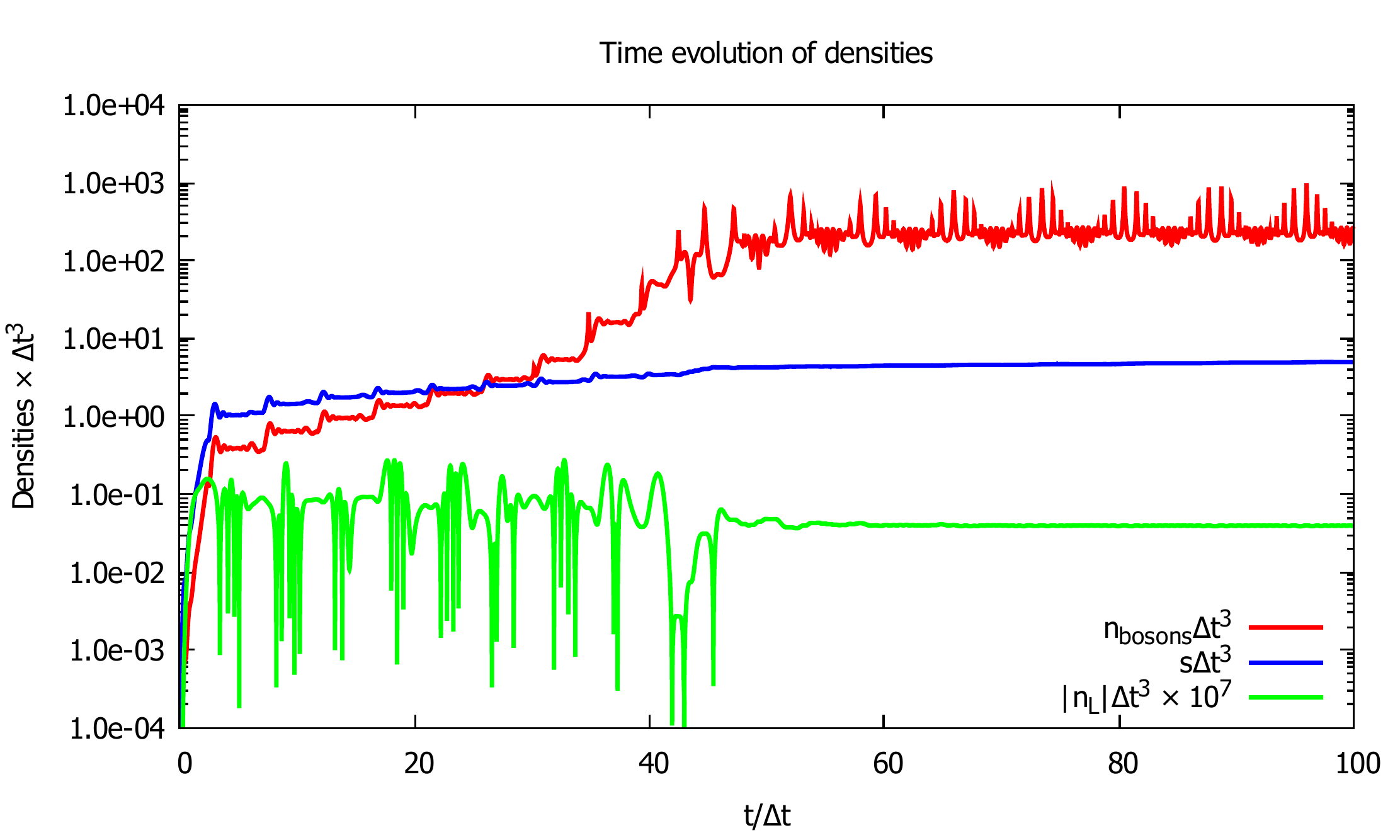}
  \caption{Numerical results with parameters $\langle h(t_0)\rangle=1.5\times 10^{14}$ GeV and $M_1=10^{15}$ GeV.  {\bf Upper:} The time evolution of the lepton-to-entropy ratio, its absolute value with the logarithmic scale and the amplitude of the Higgs background.  The final value of the lepton-to-entropy ratio at $t/\Delta t=100$ is $n_L/s=-6.54\times 10^{-10}$.  {\bf Lower:} The time evolution of the number of bosons, their corresponding entropy and lepton number in volume $\Delta t^3$.}
  \label{fig:succesful_example}
 \end{center}
\end{figure}

Next, we focus on the situation of Case B in which the entropy production occurs through the resonant particle production of the gauge bosons $W^\pm$ and $Z$ due to the oscillation of the Higgs background.   Since we assume the main production of entropy happens in this dynamics, the lepton-to-entropy ratio is fixed after all of the particle production due to the Higgs background completes.  In this case, surprisingly, enough lepton asymmetry can be generated even if the scale of the amplitude of the Higgs background is smaller than Case A.  As an example of the successful results, we show the numerical results with the parameters
\begin{equation}
 \langle h(t_0)\rangle = 1.5 \times 10^{14}\:\:{\rm GeV}, \quad
 M_1 = 10^{15} \:\:{\rm GeV}
\end{equation}
in Figure \ref{fig:succesful_example}.  The upper panel shows the time evolution of the lepton-to-entropy ratio and the amplitude of the Higgs background.  Although the aspects of the time evolutions are similar as seen in Figure \ref{fig:result_higher_amplitude}, one can see a different point that the graph of $|n_L/s|$ seems to decrease in its evolution during the flipping of the asymmetry.  The reason can be seen from the lower panel of Figure \ref{fig:succesful_example} that shows the time evolutions of the number density of bosons, entropy density, and net lepton density in the volume $\Delta t^3$.
Actually, the magnitude of the net lepton number is almost fixed except its sign.  But the entropy is generated exponentially by the parametric resonance.  As a result, $n_L/s$ reduces at the early stage, and the magnitude is fixed because the entropy production becomes steady at the later stage.  It is also interesting that the produced entropy is much smaller than the bosonic number density.

\begin{figure}[t]
 \begin{center}
  \includegraphics[scale=0.37]{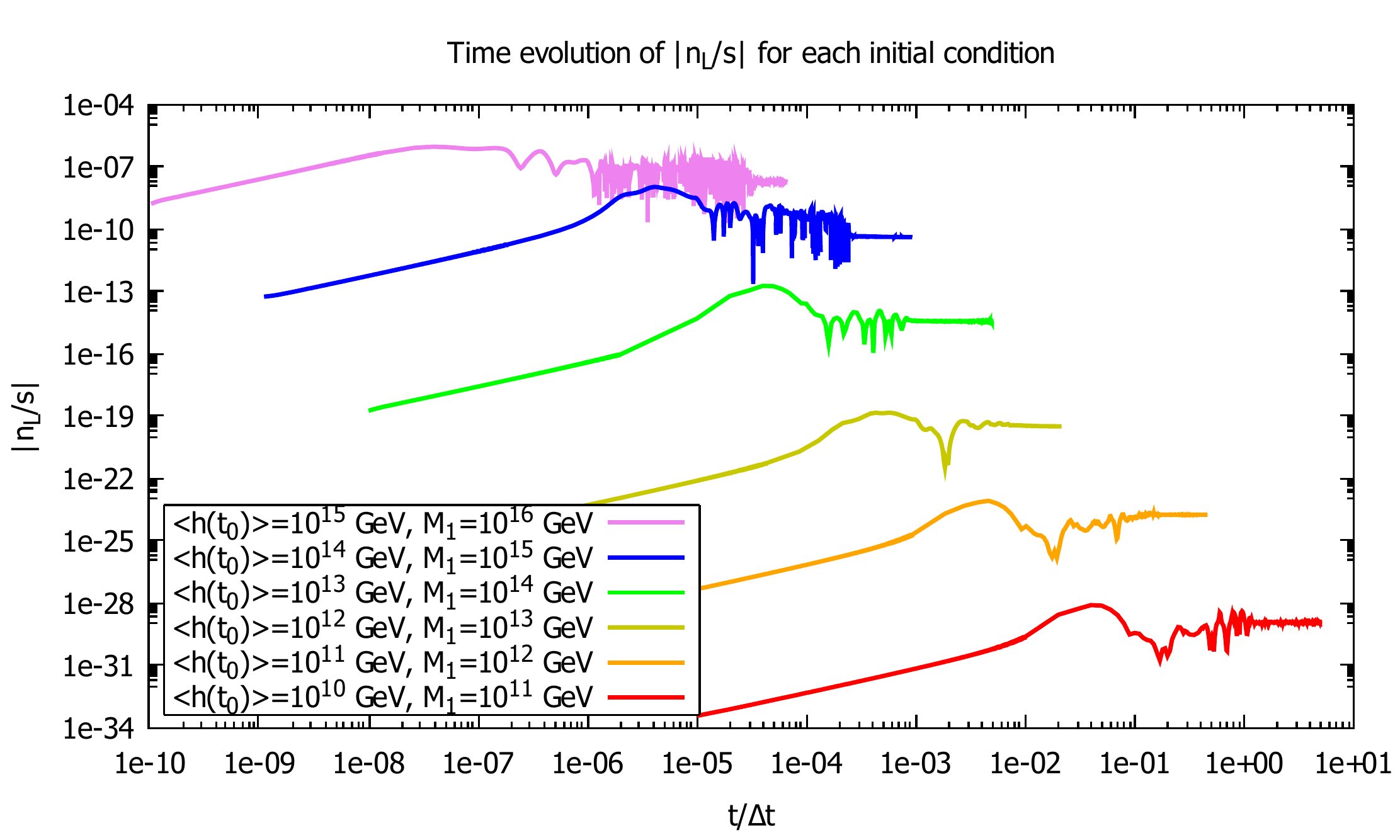}
  \caption{The time evolution of the absolute value of the lepton-to-entropy ratio comparison with several initial conditions.  The initial value of $\langle h(t_0)\rangle$ and the lightest right-handed neutrino mass $M_1$ are varied but their ratio is fixed as $\langle h(t_0)\rangle/M_1=0.1$.  The time scale $\Delta t$ is evaluated in the case of $\langle h(t_0)\rangle=10^{10}$ GeV.}
  \label{fig:comparison_asym}
 \end{center} 
\end{figure}

Finally, we show the comparison with different values for $\langle h(t_0)\rangle$ and $M_1$ in Figure \ref{fig:comparison_asym}.  According to this result, larger scale of $\langle h_(t_0)\rangle$ and $M_1$ gives larger magnitude of $n_L/s$.  Although we fix $\langle h(t_0)\rangle/M_1=0.1$ in this comparison, the case of $\langle h(t_0)\rangle/M_1<0.1$ leads smaller magnitude of $n_L/s$ during the whole time evolution.  This figure also shows that more than $10^{14}$ GeV scales for $\langle h(t_0)\rangle$ and $M_1$ are required to generate $|n_L/s|\sim 10^{-10}$ in our scenario.

\section{Conclusion and discussion} \label{sec:conclusion}
In this paper, we proposed a new leptogenesis scenario in which the lepton asymmetry is generated by the coherent oscillating Higgs background, and demonstrated that the type-I seesaw model as an illuminating example can generate enough lepton number.  Although the analytic results are not derived because of the difficulty of the analysis, we showed the numerical results with some choices of parameters.  We emphasize that in our scenario the particle production and the asymmetry generation occur simultaneously.  Hence any perturbative decay processes do not need during the lepton asymmetry generation.  This is a quite different point from the ordinary scenario.  Although we discussed the type-I seesaw model, a similar scenario is possible to be constructed by other models that include the baryon or lepton number violating interaction, C- and CP-violating parameters, and the time-varying background fields.  If such a model has the oscillating background field and baryon or lepton number violating interaction with C and CP violation, then the asymmetry could be generated.

We considered two cases of entropy production separately.  One is due to the perturbative decay of the inflation field, that is considered in the standard reheating theory.  In this situation, the typical amplitude scale of the Higgs background is required around $\langle h\rangle\sim 10^{16}$ GeV.  Another case is due to the non-perturbative decay from the oscillating Higgs background, in which the Higgs inflation model would be applicable.  An advantage compared to Case A, lower scale $\langle h\rangle\sim 10^{14}$ GeV is possible.  Note that the parameter choice in section \ref{sec:NR} is an example in whole parameter space.  For example, there exists the allowed parameter choice for $\langle h_0\rangle \gtrsim 10^{14}$ GeV in Case B if other parameters (e.g., $M_i, \theta_{ij},$ etc.) are chosen to the different values.

In both cases, the quite large initial value for the Higgs background compared to the electroweak scale is required.  In Case B, the origin of such a large value would be expected that the quantum fluctuation grows before or during inflation as similar to the chaotic inflation because the Higgs is identified to the inflaton\footnote{The chaotic-type Higgs inflation is already ruled out because the CMB spectrum requires $\lambda\sim 10^{-13}$.  The successful Higgs inflation model \cite{Bezrukov:2007ep} has the non-minimal coupling between the Higgs and gravity.  In the case of $\lambda=0.001$, the logarithmic potential is realized at $|\langle h\rangle| \gtrsim 6\times 10^{16}$ GeV.  In $|\langle h\rangle| \lesssim 6\times 10^{14}$ GeV, the Higgs potential behaves as the SM.}.  However, to apply this expectation might be not straightforward in Case A.  Once the Higgs achieves the large expectation value before or during inflation, its value can be maintained as long as the Higgs oscillation scale is smaller than the Hubble friction $\sqrt{\lambda}\langle h_0\rangle \lesssim H$.  If we consider the chaotic inflation, then $H\sim 10^{13-14}$ GeV at the end of inflation is expected.  Then, the initial Higgs background must be $\langle h_0\rangle \lesssim H_{\rm end}/\sqrt{\lambda}\sim 3\times 10^{14-15}$ GeV in order to start the Higgs oscillation after inflation.  As we have seen in (\ref{eq:lepton_not_suitable}), the lower initial value of $\langle h_0\rangle$ leads a unfavorable requirement $\Gamma_\phi \gtrsim m_\phi$ to explain the present baryon asymmetry.  Thus, if one wants to seek the origin of the large initial value $\langle h_0\rangle$ by the quantum fluctuation, additional conditions might be required to maintain the Higgs initial background in Case A, e.g., the existence of a flat-shape potential at $\langle h\rangle > \langle h_0\rangle$ or other mechanisms.

\vspace{2ex}

Because we used some approximations in our analysis to avoid complicated formulations, we need to mention the validity and the condition.

First, we constructed an effective theory in which the right-handed neutrinos do not appear.  If one wants to apply to a case that the scale of the initial amplitude of the Higgs background is larger than the mass scale of the right-handed neutrinos, a complete calculation is needed.

Secondary, we neglected the fermionic two-point functions and any correlation functions of more than two-points.  In the case that the bosonic resonant production is not relevant, the fermionic two-point function could affect the backreaction in (\ref{eq:backreaction_term}) as well as the bosonic terms.  
If the effects of the correlation functions of three and more points could be included in the analysis, the results would describe the effects of decay and scattering processes, which could provide secondary sources of the lepton asymmetry.

Thirdly, our analysis neglects the decay of the Higgs and the gauge bosons.  If those particles rapidly decay in the leptogenesis time scale, the parametric resonance that reduces the energy of the Higgs background does not occur.  For example, the decay rate of $W$ can be evaluated by
\begin{equation}
 \Gamma_W = 0.058 \cdot g_W^2m_W=0.058\cdot \frac{1}{2}g_W^3 |\langle h\rangle|
\end{equation}
where the coefficient $0.058$ is taken to satisfy the current experimental result $\Gamma_W=2$ GeV \cite{Haller:2018nnx,Zyla:2020zbs} at the electroweak scale.
Applying our parameter (\ref{eq:value_couplings}), one can obtain
\begin{equation}
 \Gamma_W = 0.0051 |\langle h\rangle|.
\end{equation}
Using the Boltzmann equation neglected the expansion effect
\begin{eqnarray}
 \partial_t n_W &=& -\Gamma_W(t)n_W
\end{eqnarray}
where $n_W$ is the number density of $W$ boson, we can estimate the decrease rate of $W$ bosons in each interval of the particle production as
\begin{eqnarray}
 \frac{n_W(t+2t_{\rm osc})}{n_W(t)}
 &=& \exp\left[-\int_t^{t+2t_{\rm osc}}dt'\:\Gamma_W(t')\right] \nonumber \\
 &=& \exp\left[-2\int_0^{\langle h_{\rm max}\rangle}dh\frac{0.0051h}{\sqrt{\frac{\lambda}{2}\left(\langle h_{\rm max}\rangle^4-h^4\right)}}\right] \nonumber \\
 &=& e^{-0.0051\pi/\sqrt{2\lambda}} \: = \: 0.70
\end{eqnarray}
where $t_{\rm osc}$ is the quarter-oscillation time defined in (\ref{eq:osc_time_scale}) and we used (\ref{eq:dot_h_relation}) as going to the second line.  Because of the decay rate of $Z$, $\Gamma_Z=2.5$ GeV, the decrease rate of $Z$ could be similar. 
Since 70\% of $W$ and $Z$ bosons could survive until the next particle production, we can expect that still the parametric resonance can work well.  If the decay of $W$ and $Z$ is taken into account, the end time of the leptogenesis would be slightly longer because the resonance  efficiency is reduced.

Finally, we neglected the spatial expanding effect in the whole calculation.  Although a realistic model must include the expansion effect, we ignored it for simplicity and to see a clear structure of the dynamics.  As we mentioned above, the Higgs background can maintain the initial value of its amplitude when the Hubble parameter $H$ is much larger than the oscillation scale of the Higgs background ($H\gg\sqrt{\lambda}|\langle h_{\rm max}\rangle|$).  The Higgs background can start to oscillate after the expansion scale becomes smaller than the oscillation scale ($H\lesssim\sqrt{\lambda}|\langle h_{\rm max}\rangle|$).  However, the expanding effect might change the whole dynamics seriously.  Since the time scale of the particle production $\Delta t$ is much smaller than the Hubble inverse $H^{-1}$ in many cases, the expanding effect at the moment of the particle production can be negligible.  But the spatial expansion makes the amplitude of the Higgs background shrink, and thus the velocity of the background at the particle production area becomes smaller.  Consequently, the amount of the produced left-handed neutrinos would be reduced, and thus there is a possibility that the lepton asymmetry would become smaller.  On the other hand, entropy production would also be reduced.  The result might strongly depend on the evolution of the Hubble parameter, i.e., the matter contents.  We leave the analysis with the expanding effect to a future work.

\section*{Acknowledgements}
This work is supported in part by the National Natural Science Foundation of China under Grants No.~11805288, No.~11875327, and No. 11905300, the China Postdoctoral Science
Foundation under Grant No. 2018M643282, the Natural Science Foundation of Guangdong Province under Grant No.~2016A030313313, the Fundamental Research Funds for the Central Universities, and the Sun Yat-Sen University Science Foundation.


\begin{thebibliography}{1}

\bibitem{Aghanim:2018eyx} 
  N.~Aghanim {\it et al.} [Planck Collaboration],
  ``Planck 2018 results. VI. Cosmological parameters,''
  arXiv:1807.06209 [astro-ph.CO].
  
\bibitem{Ade:2015xua}
P.~Ade \textit{et al.} [Planck],
``Planck 2015 results. XIII. Cosmological parameters,''
Astron. Astrophys. \textbf{594}, A13 (2016)
doi:10.1051/0004-6361/201525830
[arXiv:1502.01589 [astro-ph.CO]].

\bibitem{Cyburt:2015mya}
R.~H.~Cyburt, B.~D.~Fields, K.~A.~Olive and T.~H.~Yeh,
``Big Bang Nucleosynthesis: 2015,''
Rev. Mod. Phys. \textbf{88}, 015004 (2016)
doi:10.1103/RevModPhys.88.015004
[arXiv:1505.01076 [astro-ph.CO]].

\bibitem{Kofman:1997yn}
L.~Kofman, A.~D.~Linde and A.~A.~Starobinsky,
``Towards the theory of reheating after inflation,''
Phys. Rev. D \textbf{56}, 3258-3295 (1997)
doi:10.1103/PhysRevD.56.3258
[arXiv:hep-ph/9704452 [hep-ph]].

\bibitem{Amin:2014eta}
M.~A.~Amin, M.~P.~Hertzberg, D.~I.~Kaiser and J.~Karouby,
``Nonperturbative Dynamics Of Reheating After Inflation: A Review,''
Int. J. Mod. Phys. D \textbf{24}, 1530003 (2014)
doi:10.1142/S0218271815300037
[arXiv:1410.3808 [hep-ph]].

\bibitem{Lozanov:2019jxc}
K.~D.~Lozanov,
``Lectures on Reheating after Inflation,''
[arXiv:1907.04402 [astro-ph.CO]].

\bibitem{Enomoto:2020epjc}
S.~Enomoto, C.~Cai, Z.~H.~Yu, H.~H.~Zhang,
``Matter-antimatter asymmetry in preheating,''
AAPPS Bull. \textbf{30} (2020) no. 5, 45–48.
http://aappsbulletin.org/myboard/read.php?Board=focus\&id=125

\bibitem{Dolgov:1989us}
A.~Dolgov and D.~Kirilova,
``ON PARTICLE CREATION BY A TIME DEPENDENT SCALAR FIELD,''
Sov. J. Nucl. Phys. \textbf{51}, 172-177 (1990)
JINR-E2-89-321.

\bibitem{Traschen:1990sw}
J.~H.~Traschen and R.~H.~Brandenberger,
``Particle Production During Out-of-equilibrium Phase Transitions,''
Phys. Rev. D \textbf{42}, 2491-2504 (1990)
doi:10.1103/PhysRevD.42.2491

\bibitem{Kofman:1994rk}
L.~Kofman, A.~D.~Linde and A.~A.~Starobinsky,
``Reheating after inflation,''
Phys. Rev. Lett. \textbf{73}, 3195-3198 (1994)
doi:10.1103/PhysRevLett.73.3195
[arXiv:hep-th/9405187 [hep-th]].

\bibitem{Fukugita:1986hr}
M.~Fukugita and T.~Yanagida,
``Baryogenesis Without Grand Unification,''
Phys. Lett. B \textbf{174}, 45-47 (1986)
doi:10.1016/0370-2693(86)91126-3

\bibitem{Harvey:1990qw} 
  J.~A.~Harvey and M.~S.~Turner,
  ``Cosmological baryon and lepton number in the presence of electroweak fermion number violation,''
  Phys.\ Rev.\ D {\bf 42}, 3344 (1990).
  doi:10.1103/PhysRevD.42.3344

\bibitem{Kolb:1996jt}
E.~W.~Kolb, A.~D.~Linde and A.~Riotto,
``GUT baryogenesis after preheating,''
Phys. Rev. Lett. \textbf{77}, 4290-4293 (1996)
doi:10.1103/PhysRevLett.77.4290
[arXiv:hep-ph/9606260 [hep-ph]].

\bibitem{Kolb:1998he}
E.~W.~Kolb, A.~Riotto and I.~I.~Tkachev,
``GUT baryogenesis after preheating: Numerical study of the production and decay of X bosons,''
Phys. Lett. B \textbf{423}, 348-354 (1998)
doi:10.1016/S0370-2693(98)00134-8
[arXiv:hep-ph/9801306 [hep-ph]].

\bibitem{Giudice:1999fb}
G.~Giudice, M.~Peloso, A.~Riotto and I.~Tkachev,
``Production of massive fermions at preheating and leptogenesis,''
JHEP \textbf{08}, 014 (1999)
doi:10.1088/1126-6708/1999/08/014
[arXiv:hep-ph/9905242 [hep-ph]].

\bibitem{Dolgov:1996qq}
A.~Dolgov, K.~Freese, R.~Rangarajan and M.~Srednicki,
Phys. Rev. D \textbf{56}, 6155-6165 (1997)
doi:10.1103/PhysRevD.56.6155
[arXiv:hep-ph/9610405 [hep-ph]].

\bibitem{Kusenko:2014uta}
A.~Kusenko, K.~Schmitz and T.~T.~Yanagida,
Phys. Rev. Lett. \textbf{115}, no.1, 011302 (2015)
doi:10.1103/PhysRevLett.115.011302
[arXiv:1412.2043 [hep-ph]].

\bibitem{Funakubo:2000us}
K.~Funakubo, A.~Kakuto, S.~Otsuki and F.~Toyoda,
``Charge generation in the oscillating background,''
Prog. Theor. Phys. \textbf{105}, 773-788 (2001)
doi:10.1143/PTP.105.773
[arXiv:hep-ph/0010266 [hep-ph]].

\bibitem{Rangarajan:2001yu}
R.~Rangarajan and D.~V.~Nanopoulos,
Phys. Rev. D \textbf{64}, 063511 (2001)
doi:10.1103/PhysRevD.64.063511
[arXiv:hep-ph/0103348 [hep-ph]].

\bibitem{Enomoto:2017rvc}
S.~Enomoto and T.~Matsuda,
``Asymmetric preheating,''
Int. J. Mod. Phys. A \textbf{33}, no.25, 1850146 (2018)
doi:10.1142/S0217751X18501464
[arXiv:1707.05310 [hep-ph]].

\bibitem{Weinberg:1979sa}
  S.~Weinberg,
  ``Baryon and Lepton Nonconserving Processes,''
  Phys.\ Rev.\ Lett.\  {\bf 43} (1979) 1566.
  doi:10.1103/PhysRevLett.43.1566
  
\bibitem{Kusenko:2014lra}
A.~Kusenko, L.~Pearce and L.~Yang,
Phys. Rev. Lett. \textbf{114}, no.6, 061302 (2015)
doi:10.1103/PhysRevLett.114.061302
[arXiv:1410.0722 [hep-ph]].

\bibitem{Pearce:2015nga}
L.~Pearce, L.~Yang, A.~Kusenko and M.~Peloso,
``Leptogenesis via neutrino production during Higgs condensate relaxation,''
Phys. Rev. D \textbf{92}, no.2, 023509 (2015)
doi:10.1103/PhysRevD.92.023509
[arXiv:1505.02461 [hep-ph]].

\bibitem{Pascoli:2016gkf}
S.~Pascoli, J.~Turner and Y.~L.~Zhou,
Phys. Lett. B \textbf{780}, 313-318 (2018)
doi:10.1016/j.physletb.2018.03.011
[arXiv:1609.07969 [hep-ph]].

\bibitem{Turner:2018mwh}
J.~Turner and Y.~L.~Zhou,
JHEP \textbf{01}, 022 (2020)
doi:10.1007/JHEP01(2020)022
[arXiv:1808.00470 [hep-ph]].

\bibitem{Pascoli:2018cqk}
S.~Pascoli, J.~Turner and Y.~L.~Zhou,
Chin. Phys. C \textbf{43}, no.3, 033101 (2019)
doi:10.1088/1674-1137/43/3/033101
[arXiv:1808.00475 [hep-ph]].

\bibitem{Wu:2019ohx}
Y.~P.~Wu, L.~Yang and A.~Kusenko,
JHEP \textbf{12}, 088 (2019)
doi:10.1007/JHEP12(2019)088
[arXiv:1905.10537 [hep-ph]].

\bibitem{Lee:2020yaj}
S.~M.~Lee, K.~y.~Oda and S.~C.~Park,
[arXiv:2010.07563 [hep-ph]].

\bibitem{Casas:2001sr}
  J.~A.~Casas and A.~Ibarra,
  ``Oscillating neutrinos and $\mu \to e, \gamma$,''
  Nucl.\ Phys.\ B {\bf 618} (2001) 171
  doi:10.1016/S0550-3213(01)00475-8
  [hep-ph/0103065].

\bibitem{Bezrukov:2007ep}
F.~L.~Bezrukov and M.~Shaposhnikov,
Phys. Lett. B \textbf{659}, 703-706 (2008)
doi:10.1016/j.physletb.2007.11.072
[arXiv:0710.3755 [hep-th]].

\bibitem{Abe:2019vii}
K.~Abe \textit{et al.} [T2K],
``Constraint on the matter–antimatter symmetry-violating phase in neutrino oscillations,''
Nature \textbf{580}, no.7803, 339-344 (2020)
doi:10.1038/s41586-020-2177-0
[arXiv:1910.03887 [hep-ex]].

\bibitem{Degrassi:2012ry}
G.~Degrassi, S.~Di Vita, J.~Elias-Miro, J.~R.~Espinosa, G.~F.~Giudice, G.~Isidori and A.~Strumia,
``Higgs mass and vacuum stability in the Standard Model at NNLO,''
JHEP \textbf{08}, 098 (2012)
doi:10.1007/JHEP08(2012)098
[arXiv:1205.6497 [hep-ph]].

\bibitem{Buttazzo:2013uya}
D.~Buttazzo, G.~Degrassi, P.~P.~Giardino, G.~F.~Giudice, F.~Sala, A.~Salvio and A.~Strumia,
``Investigating the near-criticality of the Higgs boson,''
JHEP \textbf{12}, 089 (2013)
doi:10.1007/JHEP12(2013)089
[arXiv:1307.3536 [hep-ph]].

\bibitem{Haller:2018nnx}
J.~Haller, A.~Hoecker, R.~Kogler, K.~M\"onig, T.~Peiffer and J.~Stelzer,
Eur. Phys. J. C \textbf{78}, no.8, 675 (2018)
doi:10.1140/epjc/s10052-018-6131-3
[arXiv:1803.01853 [hep-ph]].

\bibitem{Zyla:2020zbs}
P.~A.~Zyla \textit{et al.} [Particle Data Group],
PTEP \textbf{2020}, no.8, 083C01 (2020)
doi:10.1093/ptep/ptaa104
  
\end{thebibliography}
\end{document}